\documentclass%
[aip,reprint,twocolumn,amsmath,amssymb]%
{revtex4-1}

\pdfoutput=1

\usepackage{graphicx}
\usepackage[colorlinks=true,allcolors=blue]{hyperref}

\newcommand{\vn}{\vec n}
\newcommand{\Rd}{a}
\newcommand{\Ea}{E_\mathrm{\Rd}}
\newcommand{\vectEa}{\vec{E}_\mathrm{\Rd}}
\newcommand{\epsilone}{\varepsilon_\mathrm{e}}
\newcommand{\epsiloni}{\varepsilon_\mathrm{i}}
\newcommand{\lambdae}{\lambda_\mathrm{e}}
\newcommand{\lambdai}{\lambda_\mathrm{i}}
\newcommand{\lambdas}{\lambda_\mathrm{s}}
\newcommand{\etae}{\eta_\mathrm{e}}
\newcommand{\etai}{\eta_\mathrm{i}}
\newcommand{\sigmas}{\sigma_{\rm s}}

\begin{document}

\title{Unsteady electrorotation of a drop in a constant electric field}

\author{Alexander N. Tyatyushkin}
 \email{tan@imec.msu.ru}
 \affiliation
 {Institute of Mechanics, Moscow State University,
  Michurinskiy Ave., 1, Moscow 119192, Russia}

\begin{abstract}
The unsteady electrorotation of a drop of a viscous weakly conducting polarizable liquid suspended in another viscous weakly conducting polarizable liquid immiscible with the former in an applied constant uniform electric field is theoretically investigated. The surface tension of the drop is regarded as sufficiently large for deformation of the drop under the action of the applied electric field and electrohydrodynamic flow can be considered small and the drop can be considered spherical in the calculation of the electric field and flow. The electric field intensity and the velocity and pressure in the electrohydrodynamic flow are sought for in the form of a series solution with time-dependent scalar, vector, and tensor coefficients for which relations are found allowing one to determine them. The coefficients are sought for in the form of asymptotic expansions over the parameter the smallness of which corresponds to the sufficiently large viscosity of the drop with respect to that of the surrounding liquid. The differential equations for the terms of the dimensionless expansions up to the first and second orders with respect to this parameter are obtained. Their stationary solutions are found in the explicit form and their stability is investigated. Up to the terms of the second order, transitions to deformational oscillations of the drop are established not to occur.
\end{abstract}


\maketitle

\section{Introduction}

Under the action of an applied electric field, electrohydrodynamic flow arises inside and outside a drop of viscous weakly conducting polarizable liquid suspended in another viscous weakly conducting polarizable liquid immiscible with the former. If the ratio of the conductivity of the drop to the conductivity of the surrounding liquid is greater than the ratio of the dielectric  permittivity of the drop to the dielectric permittivity of the surrounding liquid, the drop rotates in a constant electric field when the electric field intensity exceeds some threshold value. This phenomenon is called electrorotation. It is caused by the presence of the surface convective electric current at the interface between the liquids. Unlike the electrohydrodynamic flow, the electrorotation takes place not only for suspended drops, but also for suspended solid particles. The interest to the study of the electrohydrodynamic flows with electrorotation is caused by the possible application of this phenomenon to the flow control in various devices, in particular in microfluidics.

Experiments show that, if the electrorotation takes place, the drop deforms in a constant electric field so that it either tends to take some stationary non-axisymmetric shape \cite{Salipante&Vlahovska10,He&al} or performs deformational oscillations\cite{Salipante&Vlahovska13}. If the ratio of the viscosity of the drop to the viscosity of the surrounding liquid is not too large, hysteretic phenomena are observed when the intensity of the applied electric field varies\cite{Salipante&Vlahovska10}.

Melcher and Taylor \cite{Melcher&Taylor} investigated theoretically the electrorotation of a rigid weakly conducting polarizable infinitely long cylinder in a viscous weakly conducting polarizable liquid. Jones \cite{Jones} studied the electrorotation of a rigid weakly conducting polarizable spherical particle. The electrorotataion of rigid spherical shells \cite{Turcu}, rigid long cylinders \cite{Lemaire&Lobry}, and spheroids \cite{Cebers&al} was also studied. The electrorotation of a drop of viscous weakly conducting polarizable liquid was theoretically and experimentally investigated in Refs.  \onlinecite{Krause&Chandratreya,Ha&Yang,Sato&al},\onlinecite{Salipante&Vlahovska10,Salipante&Vlahovska13,He&al}, and  \onlinecite{Yariv&Frankel}. The influence of the non-rotational charge convection over the surface of a drop on the electrohydrodynamic flow is investigated in Refs.  \onlinecite{Shkadov&Shutov,Shutov,Lanauze&al,Das&Santillian,Tyatyushkin}.

A review of early works on the electrohydrodynamic flows, including the works devoted to the electrorotation, is given in the paper of Melcher and Taylor \cite{Melcher&Taylor}. A review of works devoted to electrorotation of rigid particles is presented in the work of Jones \cite{Jones}. Reviews of the researches devoted to electrorotation of drops can be found in the works \cite{Salipante&Vlahovska10,He&al,Salipante&Vlahovska13}. The most recent and complete review of the works devoted to both experimental and theoretical investigations of the electrorotation of drops is presented in the paper of Vlahovska \cite{Vlahovska16}.
 
The goal of this work is to investigate theoretically the unsteady electrorotation of a drop of viscous weakly conducting polarizable liquid suspended in another viscous weakly conducting polarizable liquid immiscible with the former in a constant electric field with taking into account the surface conductivity. The theory presented in Ref. \onlinecite{He&al} is supposed to be extended here in the following three directions: first, using the approximation of very viscous drops as the zeroth approximation, to investigate the asymptotic expansion over a parameter that tends to zero as the viscosity of the drop tends to infinity, second, to take into account the surface conductivity of the interface between the drop and the surrounding liquid, and third, to consider non-stationary electric fields and unsteady elctrohydrodynamic flows. The study of non-stationary solutions for stationary external conditions allows one to investigate the stability of the stationary solutions and to ascertain the possibility of the solutions with a transition to steady-state oscillations. The transition to oscillations and chaotic behavior was studied experimentally and theoretically in Ref. \onlinecite{Salipante&Vlahovska13}. The theory presented in Ref. \onlinecite{Salipante&Vlahovska13} is based on the assumption that the drop has the shape of an oblate spheroid with fixed aspect ratio. In the given work, the shape of the drop is supposed to be determined by equations obtained with the help of asymptotic methods.

In Section II, the problem is set, the governing equations and boundary conditions are written down, and the used approximations are discussed. In Section III, the solution is written down in the form of series with time-dependent scalar, vector, and tensor coefficients, the dimensionless coefficients are sought for in the form of asymptotic expansions over a parameter smallness of which corresponds to large viscosity of the drop with respect to that of the surrounding liquid, and the expressions and differential equations for the terms of those expansions up to the first and second orders are written down and analyzed. The obtained results are summarized and discussed in Section IV. The details of the calculations are presented in Appendix A and Appendix~B.

\section{Setting of the problem}

\subsection{Object of investigation}

\begin{figure}
\includegraphics{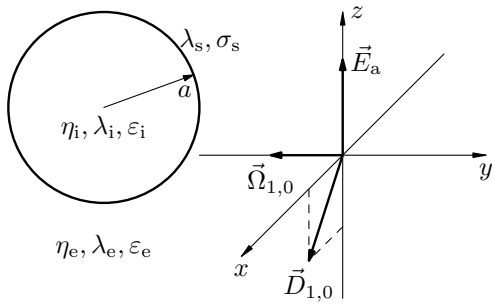}
\caption{
Setting of the problem
}
\label{f1}
\end{figure}

Consider a drop of incompressible liquid in an applied constant uniform electric field with intensity $\vectEa$ (see Fig.~\ref{f1}). The viscosity, electric conductivity, and dielectric permittivity of the liquid inside the drop are $\etai$, $\lambdai$, and $\epsiloni$. The radius of the drop in the undeformed state is $\Rd$. The drop is suspended in an incompressible liquid with the viscosity $\etae$, electric conductivity $\lambdae$, and dielectric permittivity $\epsilone$. The surface tension and surface electric conductivity of the interface between the liquids are $\sigmas$ and $\lambdas$. The liquids are regarded as sufficiently viscous so that the small Reynolds number approximation is valid and the conductivities are so small that the electrohydrodynamic approximation is valid (see Ref. \onlinecite{Melcher&Taylor}).

\subsection{System of equations}

\newcommand{\js}{j_\mathrm{s}}
\newcommand{\e}[1]{\vec e_#1}
\newcommand{\vrs}{\vec r_{\rm s}}

\newcommand{\tsigmae}{\hat\sigma_{\rm e}}
\newcommand{\tsigmah}{\hat\sigma_{\rm v}}
\newcommand{\ac}{\mathcal{H}}

In order to find the drop shape, electric field intensity, $\vec{E}$, velocity, $\vec{v}$, and pressure, $p$, as functions of the radius-vector, $\vec{r}$, and time, $t$, under the above made assumptions, the system of equations of electrohydrodynamics (see Ref. \onlinecite{Melcher&Taylor}) written down for electric fields in the quasi-stationary approximation and for flows in the quasi-steady approximation is used. This system consists of the continuity equation for an incompressible liquid
\begin{equation}
\nabla\cdot\vec v = 0
,
\label{de}
\end{equation}
the motion equation in the small Reynolds number and quasi-steady flow approximations
\begin{equation}
\nabla\cdot
\left(-p \hat I + \tsigmae + \tsigmah\right)
=
0
,
\label{me0}
\end{equation}
the Maxwell's equations in the electrohyrodynamic and quasi-stationary field approximations
\begin{equation}
\nabla \cdot \vec D = 0
,\qquad
\nabla \times \vec E = 0
,
\label{Me}
\end{equation}
the electric charge conservation law
\begin{equation}
\nabla \cdot \vec j = 0
,
\label{j}
\end{equation}
and the constitutive relations 
\begin{eqnarray}
\vec j &=& \lambda \vec E
,
\\*
\vec D &=& \varepsilon \vec E
.
\label{DE}
\end{eqnarray}
Here, the formulas are written down for Gaussian system of units, $\cdot$ and $\times$ denote the scalar and vector products, $\vec{D}$ is the electric induction, $\vec{j}$ is the volume density of the electric current, $\eta=\etai$, $\lambda=\lambdai$, and  $\varepsilon=\epsiloni$ inside the drop, $\eta=\etae$, $\lambda=\lambdae$, and  $\varepsilon=\epsilone$ outside it, $\nabla$ is the nabla operator, $\hat{I}$ is the identity tensor, $\tsigmae$ and $\tsigmah$ are the tensors of electric and viscous stresses expressed as follows:
\begin{eqnarray}
\tsigmae
&=&
\frac{1}{4\pi} \vec D \vec E
-
\frac{1}{8\pi} \vec D \cdot \vec E \hat I
,
\label{sigmae}
\\*
\tsigmah
&=&
2 \eta \left(\nabla \vec v\right)^\mathrm{S}
,
\label{sigmah}
\end{eqnarray}
where $\vec{A}\vec{B}$ denotes the dyadic product of the vectors $\vec{A}$ and $\vec{B}$, $\nabla\vec{f}$ denotes the dyadic product of the nabla operator and the vector field $\vec{f}=\vec{f}(\vec{r})$, $\hat{T}^\mathrm{S}$ denotes the symmetric part of the tensor $\hat{T}$. Since the electric conductivity and dielectric permittivity are uniform, the electric charge conservation law, Eq.~(\ref{j}), follows from Maxwell's equations for electric induction Eq.~(\ref{Me}).

Using the continuity equation (\ref{de}) and Maxwell's equations (\ref{Me}) and taking into account that the electric conductivity and dielectric permittivity are uniform, the motion equation (\ref{me0}) can be rewritten in the form of the Navier--Stokes equation in the small Reynolds number approximation \cite{Melcher&Taylor} 
\begin{equation}
-\nabla p + \eta \Delta \vec v
=
0
,
\label{me}
\end{equation}
where $\Delta$ is the Laplacian.

\subsection{Boundary conditions} 

The boundary conditions on the interface between the liquids include the impenetrability condition 
\begin{equation}
\left.\vec v\right|_{\rm e}\cdot\vec{n}
=
\left.\vec v\right|_{\rm i}\cdot\vec{n}
=
{v}_{\mathrm{s}n}
,
\label{BC:vn}
\end{equation}
the no-slip condition
\begin{equation}
\left[\vec v\right]_{\rm s}\times\vec n=0
,
\label{BC:vt}
\end{equation}
the conditions for the jumps of the normal and tangential components of the stress vector $\vec\sigma_n=\vec{n}\cdot\left(-p\hat{I}+\tsigmae+\tsigmah\right)$
\begin{eqnarray}
\vec{n}\cdot
\left[
 -p \hat I + \tsigmae + \tsigmah
\right]_{\rm s}
\cdot\vec{n}
&=&
-
\sigma_{\rm s} \ac
,
\label{BC:sn}
\\*
\vec{n}\cdot
\left[
 \tsigmae + \tsigmah
\right]_{\rm s}
\times\vec{n}
&=&
0
,
\label{BC:st}
\end{eqnarray}
the continuity condition for the tangential component of the intensity of the electric field
\begin{eqnarray}
\left[\vec E\right]_{\rm s}\times\vec{n}=0
,
\label{BC:E}
\end{eqnarray}
and the conditions for the jump of the normal component of the density of the electric current 
\begin{equation}
\dfrac{\partial q_\mathrm{s}}{\partial t}
=
q_\mathrm{s}\mathcal{H}{v}_{\mathrm{s}n}
-
\left[\vec{j}\right]_{\rm s}\cdot\vec{n}
-
\nabla_\alpha j_\mathrm{s}^\alpha
=
q_\mathrm{s}\mathcal{H}{v}_{\mathrm{s}n}
-
\left[\vec{j}\right]_{\rm s}\cdot\vec{n}
-
\nabla_\mathrm{s}\cdot\vec{j}_\mathrm{s}
.
\label{BC:j-0}
\end{equation}
Here, $\left.A\right|_{\textrm{i}}$ and $\left.A\right|_{\textrm{e}}$ denote the values of the quantity $A$ on the interface between the liquids approached from inside and outside the drop, respectively,
$[A]_{\rm s}=\left.A\right|_{\rm e}-\left.A\right|_{\rm i}$
denotes the jump of the quantity $A$ at the interface when moving from the inside to the outside, $\nabla_\mathrm{s}$ denotes the surface nabla operator, $\vec{n}$ is the external normal unit vector at a given point of the interface, ${v}_{\mathrm{s}n}$ is the normal component of the velocity of the surface of the drop at a given point, $\mathcal{H}$ is the mean curvature at a given point of the surface of the drop, $q_\mathrm{s}$ is the density of the surface electric charge, determined by the condition for the jump of the the normal component of the electric induction 
\begin{equation}
\left[\vec{D}\right]_{\rm s}\cdot\vec{n}
=
4\pi q_\mathrm{s}
,
\label{BC:D}
\end{equation}
$\vec{j}_\mathrm{s}$ is the density of the total surface electric current. Note that definition of the mean curvature used in the present work is such that it takes negative values on the surface of a convex domain.

The continuity conditions for the tangential components of the velocity and electric field intensity allow determining the following vector fields defined on the surface of the drop:
\begin{equation}
\vec{v}_{\mathrm{s}\tau}
=
\vec{n}\times
\left(\left.\vec{v}\right|_\mathrm{i}\times\vec{n}\right)
=
\vec{n}\times
\left(\left.\vec{v}\right|_\mathrm{e}\times\vec{n}\right)
,
\label{d:vs}
\end{equation}
\begin{equation}
\vec{E}_\mathrm{s}
=
\vec{n}\times
\left(\left.\vec{E}\right|_\mathrm{i}\times\vec{n}\right)
=
\vec{n}\times
\left(\left.\vec{E}\right|_\mathrm{e}\times\vec{n}\right)
.
\label{d:Es}
\end{equation}
The density of the total surface electric current is determined with the help of these functions as follows:
\begin{equation}
\vec{j}_\mathrm{s}
=
\lambda_\mathrm{s} \vec{E}_\mathrm{s}
+
q_\mathrm{s} \vec{v}_{\mathrm{s}\tau}
.
\label{js}
\end{equation}
Substituting Eq.~(\ref{js}) into Eq.~(\ref{BC:j-0}), one obtains
\begin{equation}
\dfrac{\partial q_\mathrm{s}}{\partial t}
=
q_\mathrm{s}\mathcal{H}{v}_{\mathrm{s}n}
-
\left[\vec{j}\right]_{\rm s}\cdot\vec{n}
-
\nabla_\mathrm{s}\cdot
\left(\lambda_\mathrm{s} \vec{E}_\mathrm{s}\right)
-
\nabla_\mathrm{s}\cdot
\left(q_\mathrm{s} \vec{v}_{\mathrm{s}\tau}\right)
.
\label{BC:j}
\end{equation}

Using the boundary conditions (\ref{BC:E}) and (\ref{BC:D}) and the constitutive relations, Eq.~(\ref{DE}), one can transform the conditions for the stress vector, Eq.~(\ref{BC:sn}) and Eq.~(\ref{BC:st}), into the form  
\begin{multline}
-
\left[p\right]_\mathrm{s}
+
\dfrac{1}{8\pi} \epsilone
\left(\left.\vec{E}\right|_\mathrm{e}\cdot\vec{n}\right)^2
-
\dfrac{1}{8\pi} \epsiloni
\left(\left.\vec{E}\right|_\mathrm{i}\cdot\vec{n}\right)^2
\\
-
\dfrac{1}{8\pi}\left(\epsilone-\epsiloni\right)
\left|\left.\vec{E}\right|_\mathrm{i}\times\vec{n}\right|^2
+
2\vec{n}\cdot
\left[\eta\nabla\vec{v}\right]_\mathrm{s}
\cdot
\vec{n}
=
-
\sigma_\mathrm{s} \ac
,
\end{multline}
\begin{equation}
\vec{n}\times
\left(
 \left[\eta\nabla\vec{v}\right]_\mathrm{s}\cdot\vec{n}
 +
 \vec{n}\cdot\left[\eta\nabla\vec{v}\right]_\mathrm{s}
\right)
=
-
q_\mathrm{s}\;
\vec{n}
\times
\left.\vec{E}\right|_\mathrm{i}
.
\end{equation}
Using the introduced notations (\ref{d:vs}) and (\ref{d:Es}), one obtains 
\begin{multline}
-
\left[p\right]_\mathrm{s}
+
\dfrac{1}{8\pi} \epsilone
\left(\left.\vec{E}\right|_\mathrm{e}\cdot\vec{n}\right)^2
-
\dfrac{1}{8\pi} \epsiloni
\left(\left.\vec{E}\right|_\mathrm{i}\cdot\vec{n}\right)^2
\\
-
\dfrac{1}{8\pi}\left(\epsilone-\epsiloni\right)
{E}_\mathrm{s}^2
+
2\vec{n}\cdot
\left[\eta\nabla\vec{v}\right]_\mathrm{s}
\cdot
\vec{n}
=
-
\sigma_\mathrm{s} \ac
,
\label{bc:sn}
\end{multline}
\begin{equation}
-
\left\{
 \vec{n}\times
 \left(
  \left[\eta\nabla\vec{v}\right]_\mathrm{s}\cdot\vec{n}
  +
  \vec{n}\cdot\left[\eta\nabla\vec{v}\right]_\mathrm{s}
 \right)
\right\}
\times
\vec{n}
=
q_\mathrm{s}\vec{E}_\mathrm{s}
\end{equation}
or
\begin{equation}
-
\left(
 \vec{n}\times
 \left\{
  \left[
   2\eta\left(\nabla\vec{v}\right)^\mathrm{S}
  \right]_\mathrm{s}
  \cdot\vec{n}
 \right\}
\right)
\times
\vec{n}
=
q_\mathrm{s}\vec{E}_\mathrm{s}
\label{bc:st}
.
\end{equation}

There is the jump of the tangential component of the viscous stress vector on the surface of the drop in left part of the boundary condition (\ref{bc:st}). Thus, the boundary condition (\ref{bc:st}) describes the action of the Coulomb force $q_\mathrm{s}\vec{E}_\mathrm{s}$, which is a surface tangential force.

The boundary conditions at infinity have the form
\begin{align}
\vec v &\to 0 &\quad\text{at}\quad r &\to \infty
,
\label{BC:inf:v}
\\*
p &\to p_\infty
&\quad\text{at}\quad r &\to \infty
,
\label{BC:inf:p}
\\*
\vec E &\to \vectEa
&\quad\text{at}\quad r &\to \infty
,
\label{BC:inf:E}
\end{align}
where $p_\infty$ is the pressure at infinity.

Besides, $\vec{v}(\vec{r},t)$, $p(\vec{r},t)$, and $\vec{E}(\vec{r},t)$ should be bounded for all the bounded values of $\vec{r}$.

\subsection{Small deformation approximation}

If the following condition is satisfied
\begin{equation}
\max(\epsiloni,\epsilone)\dfrac{\Ea^2\Rd}{\sigmas}
=
\max\left(\dfrac{\epsiloni}{\epsilone},1\right) \mathsf{Ca}
\ll 
1
,
\label{approx}
\end{equation}
where $\mathsf{Ca}=\etae{V}_\mathrm{c}/\sigmas$ is the capillary number with characteristic velocity, ${V}_\mathrm{c}=\epsilone\Ea^2\Rd/\etae$, for electrohydrodynamic flows, the drop experiences small deformation under the action of the electric field and flow, i.e., its surface is almost spherical. 
In this case, in order to find the shape of the deformed drop in the first order approximation with respect to the dimensionless small parameters ${\epsiloni\Ea^2\Rd/\sigmas}$ and ${\epsilone\Ea^2\Rd/\sigmas}$, the small deformation approximation can be used. Within this approximation, the system of equations (\ref{de}), (\ref{Me})--(\ref{DE}), and (\ref{me}) is solved for the spherical drop with the use of the boundary conditions (\ref{BC:vn})--(\ref{BC:vt}), (\ref{BC:E})--(\ref{BC:j}), and (\ref{bc:st}) [i.e., without using the boundary condition (\ref{bc:sn}) following from the condition for the normal component of the stress vector, Eq.~(\ref{BC:sn})]. Then the found solution is used for the calculation of the jump of the normal component of the stress vector and, with the use of the boundary condition (\ref{bc:sn}), of the mean curvature $\ac$. Then, with the use of the calculated mean curvature, the shape of the deformed drop is found.

Let the surface of the drop be given by the following equation:
\begin{equation}
r
=
\left|\vec{r}\right|
=
\Rd
+
h\left(\dfrac{\vec{r}}{{r}},t\right)
.
\label{eq:S}
\end{equation}
The function $h\left(\vec{r}/{r},t\right)$ can be represented in the form 
\begin{equation}
h\left(\dfrac{\vec{r}}{{r}},t\right)
=
\sum\limits_{n=2}^\infty
 \hat{H}_n\stackrel{n}{\cdot}\dfrac{\vec{r}\,^n}{{r}^n}
,
\label{h}
\end{equation}
which excludes displacements of the drop by some vector without variation of its shape. Here, $\hat{H}_n=\hat{H}_n(t)$ ($n>1$) are some tensors each of which is an arbitrary irreducible tensor of $n$th order, i.e., a tensor of $n$th order symmetric with respect to any pair of indices and such that its contraction with the identity tensor over any pair of indices is equal to zero. Here and in what follows, $\vec{b}\,^n$ denotes $n$th dyadic degree of the vector $\vec{b}$ [i.e., the $(n-1)$-multiple dyadic product of the vector $\vec{b}$ by itself], $\stackrel{n}{\cdot}$ denotes $n$-multiple contraction of tensors over the adjacent indices. The condition for the small deformation approximation to be valid has the form
\begin{equation}
\max \left|h\left(\dfrac{\vec{r}}{{r}},t\right)\right|
\ll
\Rd
.
\label{cond:sda}
\end{equation}
Note that, in this representation, the the volume bounded by the surface, Eq.~(\ref{eq:S}), is equal to the volume of the undeformed drop, $4\pi\Rd^3/3$, within the small deformation approximation. Besides, in this approximation,
\begin{equation}
v_{\mathrm{s}n}
=
\sum\limits_{n=2}^\infty
 \dfrac{\mathrm{d}\hat{H}_n}{\mathrm{d}t}\stackrel{n}{\cdot}\vec{n}\,^n
.
\label{vsn-Hn}
\end{equation}
Using the condition for the small deformation approximation (\ref{cond:sda}), one may neglect the terms of higher orders with respect to function ${h}\left(\vec{r}/{r},t\right)$ and its derivatives in the calculation of the mean curvature, which yields
\begin{equation}
\mathcal{H}
=
-
\dfrac{2}{\Rd}
+
\dfrac{1}{\Rd^2}
\left[
 2 h
 -
 2\vec{n}_0\cdot\dfrac{\partial h}{\partial\vec{n}_0}
 +
 \left(\hat{I}-\vec{n}_0\vec{n}_0\right)\stackrel{2}{\cdot}
 \dfrac{\partial^2 h}{\partial{\vec{n}_0}^2}
\right]
,
\end{equation}
where $\vec{n}_0=\vec{r}/{r}$, $\partial^n{f}/\partial\vec{x}^n$ denotes the tensor that is the result of $n$-multiple differentiation of the function $f=f\left(\vec{x}\right)$ (e.g., $\nabla{f\left(\vec{r}\right)}=\partial{f}/\partial\vec{r}$). Using the representation for the function $h=h(\vec{n}_0,t)$,  Eq.~(\ref{h}), one obtains
\begin{equation}
\mathcal{H}
=
-
\dfrac{2}{\Rd}
-
\dfrac{1}{\Rd^2}
\sum\limits_{n=2}^\infty
 (n-1)(n+2)\hat{H}_n\stackrel{n}{\cdot}\dfrac{\vec{r}\,^{n}}{r^n}
.
\label{H}
\end{equation}

\section{Solution}
\subsection{Electric field}

The intensity of the electric field is sought for in the form
\begin{equation}
\vec E
=
-
\nabla\varphi
,
\label{E}
\end{equation}
where 
\begin{equation}
\varphi
=
\begin{cases}
-
\vectEa\cdot\vec{r}
+
\displaystyle
\sum\limits_{n=1}^\infty
 \hat{D}_n \stackrel{n}{\cdot} \dfrac{\vec{r}\,^n}{r^{2n+1}}
,\quad
r>\Rd
,
\\
-
\vectEa\cdot\vec{r}
+
\displaystyle
\sum\limits_{n=1}^\infty
 \hat{D}_n \stackrel{n}{\cdot} \dfrac{\vec{r}\,^n}{\Rd^{2n+1}}
,\quad
r\leqslant\Rd
.
\end{cases}
\label{phi}
\end{equation}
Here, $\hat{D}_n=\hat{D}_n(t)$ is the tensor of $2^n$th-pole electric moment which is either some vector for $n=1$ or, for $n>1$, some irreducible tensor of $n$th order. Thus, the electric field is determined by a set of tensor functions $\hat{D}_n=\hat{D}_n(t)$ ($n=1,2,\dots$). For a field given in this form, Maxwell's equations (\ref{Me}), the continuity condition of the tangential component of the intensity of the electric field, Eq.~(\ref{BC:E}), the boundary condition for the electric field at infinity, (\ref{BC:inf:E}), and the conditions of boundedness of $\vec{E}$, $\vec{D}$, and $\vec{j}$ are automatically satisfied.

Using Eq.~(\ref{phi}), the boundary conditions (\ref{d:Es}) and (\ref{BC:D}), the identity
\begin{equation}
\nabla\left[\hat{D}_n \stackrel{n}{\cdot} \vec{r}\,^n\right]
=
n \hat{D}_n \stackrel{n-1}{\cdot} \vec{r}\,^{n-1}
,
\end{equation}
and the fact that $\vec{r}=\Rd\vec{n}$ on the surface of the drop in the small deformation approximation, one obtains 
\begin{multline}
\vec{E}_\mathrm{s}
=
\vn\times\left(\vectEa\times\vn\right)
\\
-
\sum\limits_{n=1}^\infty
 n \Rd^{-n-2}
 \vn\times
 \left\{
  \left[\hat{D}_n\stackrel{n-1}{\cdot}\vec{n}\,^{n-1}\right]\times\vn
 \right\}
\label{Es}
,
\end{multline}
\begin{multline}
q_\mathrm{s}
=
\dfrac{1}{4\pi}
\left(\epsilone - \epsiloni\right)
\vectEa\cdot\vn
\\
+
\dfrac{1}{4\pi}
\sum\limits_{n=1}^\infty
 \Rd^{-n-2} \left[(n+1)\epsilone + \epsiloni n\right]
 \hat{D}_n\stackrel{n}{\cdot}\vec{n}\,^n
.
\label{qs}
\end{multline}
Using Eq.~(\ref{phi}), one also obtains
\begin{multline}
\left[\vec{j}\right]_\mathrm{s}\cdot\vn
=
\left(\lambdae - \lambdai\right)
\vectEa\cdot\vn
\\
+
\sum\limits_{n=1}^\infty
 \Rd^{-n-2} \left[(n+1)\lambdae + \lambdai n\right]
 \hat{D}_n\stackrel{n}{\cdot}\vec{n}\,^n
.
\label{[j]}
\end{multline}

Calculation of the surface divergence of the surface electric conductivity current yields
\begin{equation}
\nabla_\mathrm{s}\cdot\left(\lambdas \vec{E}_\mathrm{s}\right)
=
-
\dfrac{2}{\Rd}\lambdas\vectEa\cdot\vec{n} 
+
\lambdas
\sum\limits_{n=1}^\infty
 n(n+1)\Rd^{-n-3}\hat{D}_n \stackrel{n}{\cdot} \vec{n}\,^n
.
\label{divlambdasEs}
\end{equation}

\subsection{Electrohydrodynamic flow}

The velocity and pressure in the flow caused by the tangential surface force $q_\mathrm{s}\vec{E}_\mathrm{s}$ are determined by the continuity equation (\ref{de}) and Navier--Stokes equations in the low Reynolds number approximation, Eq.~(\ref{me}), with the impenetrability and no-slip boundary conditions, Eqs.~(\ref{BC:vn})--(\ref{BC:vt}), the conditions for tangential stresses, Eq.~(\ref{bc:st}), on the surface of the drop, the boundary conditions at infinity for the velocity and pressure, Eqs.~(\ref{BC:inf:v})--(\ref{BC:inf:p}), and the conditions of boundedness for the velocity and pressure. 

The solution is sought for in the form which can be constructed from general Lamb's solution (cf. Ref. \onlinecite{Lamb}, art. 336) for a quasi-steady flow (see \ref{A1})
\begin{equation}
\vec{v}
=
\begin{cases}
\vec{v}_\mathrm{e}
,\quad
r>\Rd
,
\\
\vec{v}_\mathrm{s}
,\quad
r=\Rd
,
\\
\vec{v}_\mathrm{i}
,\quad
r<\Rd
,
\end{cases}
\quad
p
=
\begin{cases}
{p}_\mathrm{e}
,\quad
r>\Rd
,
\\
{p}_\mathrm{i}
,\quad
r<\Rd
,
\end{cases}
\end{equation}
where $\vec{v}_\mathrm{e}$, $\vec{v}_\mathrm{s}$, $\vec{v}_\mathrm{i}$, ${p}_\mathrm{e}$, and ${p}_\mathrm{i}$ are expressed in terms of $\hat\Omega_{n}=\hat\Omega_{n}(t)$, $\hat{B}_{n}=\hat{B}_{n}(t)$, $\hat{C}_{n}=\hat{C}_{n}(t)$, and $p_0=p_0(t)$, Eqs. (\ref{ve})--(\ref{pi}). Here, $\hat\Omega_{n}=\hat\Omega_{n}(t)$, $\hat{B}_{n}=\hat{B}_{n}(t)$, $\hat{C}_{n}=\hat{C}_{n}(t)$ are some tensors depending on the time each of which is either an arbitrary vector for $n=1$ or an arbitrary irreducible tensor for $n>1$, $p_0=p_0(t)$ is the pressure at the center of the drop, which is to be found. 

It follows from Eq.~(\ref{vs}) that
\begin{multline}
\vec{v}_{\mathrm{s}\tau}
=
\Rd
\sum\limits_{n=1}^\infty
 n
 \left[
  \hat\Omega_n
  \stackrel{n-1}{\cdot} \vec{n}\,^{n-1}
 \right]
 \times\vec{n}
\\
+
\sum\limits_{n=1}^\infty
 2
 \vec{n}\times
 \left\{
  \left[\hat{B}_n \stackrel{n-1}{\cdot} \vec{n}\,^{n-1}\right]
  \times\vec{n}
 \right\}
,
\end{multline}
\begin{equation}
{v}_{\mathrm{s}n}
=
\sum\limits_{n=1}^\infty
 2(n+1) \hat{C}_n \stackrel{n}{\cdot} \vec{n}\,^{n}
,
\label{vsn}
\end{equation}
i.e., according to Eq.~(\ref{vsn-Hn}), 
\begin{equation}
\hat{C}_n
=
\dfrac{1}{2(n+1)}\dfrac{\mathrm{d}\hat{H}_n}{\mathrm{d}t}
.
\label{Cn-0}
\end{equation}
Thus, in order to solve the problem set above, the unknown vector and tensor functions $\hat{D}_n(t)$, $\hat\Omega_n(t)$, $\hat{B}_n(t)$, and $\hat{H}_n(t)$ as well as the scalar function $p_0(t)$ should be found with the use of the remaining boundary conditions (\ref{BC:j}), (\ref{bc:sn}), and (\ref{bc:st}).

\subsection{Asymptotic expansion}

Using $\Rd$, $\tau_\mathrm{v}=\left(\etae+\etai\right)\Rd/\sigmas$, $\Ea$, ${V}_\mathrm{c}=\epsilone\Ea^2\Rd/\etae$, and ${p}_\mathrm{c}=\epsilone\Ea^2$ as the characteristic length, time, electric field intensity, velocity, and pressure, the dimensionless vector, tensor, and scalar functions, $\hat{D}_n^*(t^*)$, $\hat\Omega_n^*(t^*)$, $\hat{B}_n^*(t^*)$, $\hat{H}_n^*(t^*)$, and ${p}_0^*(t^*)$, can be introduced as follows in order to adimensionalize the equations that determine them:
\begin{multline}
\hat{D}_n(t)
=
\Rd^{n+2}\Ea \hat{D}_n^*(t^*)
,\quad
\hat\Omega_n(t)
=
\dfrac{\epsilone\Ea^2}{4\pi\etae} 
\hat\Omega^*_n(t^*)
,
\\*
\hat{B}_n(t)
=
\dfrac{\epsilone\Ea^2\Rd}{8\pi\etae} 
\hat{B}_n^*(t^*)
,\quad
\hat{H}_n(t)
=
\dfrac{\etae+\etai}{\etae}
\dfrac{\epsilone\Ea^2\Rd^2}{4\pi\sigmas}
\hat{H}_n^*(t^*)
,
\\*
p_0
=
\dfrac{\epsilone\Ea^2}{4\pi}p_0^*
+
p_\infty
+
\dfrac{2\sigmas}{\Rd}
,\qquad
t
=
\dfrac{2\left(\etae+\etai\right)\Rd}{\sigmas}t^*
.
\label{adim}
\end{multline}
The adimensionalized equations contain the following dimensionless vector and scalar parameters:
\begin{multline}
\vec{k}
=
\dfrac{\vectEa}{\Ea}
,\ 
\lambdai^*
=
\dfrac{\lambdai}{\lambdae}
,\ 
\lambdas^*
=
\dfrac{\lambdas}{\Rd\lambdae}
,\ 
\epsiloni^*
=
\dfrac{\epsiloni}{\epsilone}
,\ 
\eta^*
=
\dfrac{\etae}{\etae+\etai} 
,\\
\tau^*
=
\dfrac{\epsilone^2\Ea^2}{16\pi^2\lambdae\etae} 
,\ 
\omega^*
=
\dfrac
 {\epsilone\sigmas}
 {8\pi\lambdae\left(\etae+\etai\right)\Rd}
.
\label{param*}
\end{multline}
Note that
\begin{equation}
\tau^*
=
\dfrac{{V}_\mathrm{c}}{4\pi\Rd}
\tau_\mathrm{e}
,\quad
\omega^*
=
\dfrac{\tau_\mathrm{e}}{2\tau_\mathrm{v}}
,\quad
\tau_\mathrm{e}
=
\dfrac{\epsilone}{4\pi\lambdae} 
,
\label{tau*,omega*}
\end{equation}
where $\tau_\mathrm{e}$ is the charge relaxation time in the surrounding liquid \cite{Melcher&Taylor}, i.e., $\tau^*$ is the ratio of the charge relaxation time to the characteristic time of charge convection and $\omega^*$ is the ratio of the charge relaxation time to the time of viscous relaxation of the drop shape \cite{Salipante&Vlahovska10}, $\tau_\mathrm{v}$.

The dimensionless functions $\hat{D}_n^*=\hat{D}_n^*(t^*)$, $\hat\Omega_n^*=\hat\Omega_n^*(t^*)$, $\hat{B}_n^*=\hat{B}_n^*(t^*)$, $\hat{H}_n^*=\hat{H}_n^*(t^*)$, and ${p}_0^*={p}_0^*(t^*)$ are sought for in the form of the following asymptotic expansions over the parameter $\eta^*$:
\begin{multline}
\hat\Omega_n^* 
\sim
\sum_{j=0}^\infty \eta^{*j} \hat\Omega_{n,j}^*
,\quad
\hat{D}_n^* 
\sim
\sum_{j=0}^\infty \eta^{*j} D_{n,j}^*
,
\\*
\hat{B}_n^* 
\sim
\sum_{j=0}^\infty \eta^{*j} B_{n,j}^*
,
\quad
\hat{H}_n^* 
\sim
\sum_{j=0}^\infty \eta^{*j} H_{n,j}^*
,
\\*
{p}_0^* 
\sim
\sum_{j=0}^\infty \eta^{*j} {p}_{0j}^*
,\qquad
\eta^*\to0
.
\label{asym}
\end{multline}
Note that $\eta^*\to0$ corresponds to $\etai/\etae\to\infty$. Thus, the limit case $\eta^*\to0$, i.e., the zeroth order approximation, in this problem corresponds to a rigid spherical particle in a viscous liquid.

\subsection{Zeroth order approximation and corrections of the  first order}

Using the boundary conditions (\ref{BC:j}), (\ref{bc:sn}), and (\ref{bc:st}) in the dimensionless form, one obtains the following expressions and relations for $\hat{D}_{n,0}^*$, $\hat\Omega_{n,0}^*$, $\hat{H}_{n,0}^*$, $\hat{B}_{n,0}^*$, $p_{0,0}^*$, $\hat\Omega_{n,1}^*$ ($n\neq1$), $\hat{H}_{n,1}^*$, and $\hat{B}_{n,1}^*$:
\begin{equation}
\hat\Omega_{n,0}^*
=
0
,\quad
\hat{D}_{n,0}^*
=
0
,
\quad
n\neq1
,
\label{Omegan0,Dn0}
\end{equation}
\begin{equation}
\hat{H}_{n,0}^*
=
0
,\quad
\hat{B}_{n,0}^*
=
0
,
\label{Hn0,Bn0}
\end{equation}
\begin{equation}
\vec\Omega_{1,0}^*
=
-
\dfrac{1}{2}\vec{k}\times\vec{D}_{1,0}^*
,
\label{Omega10}
\end{equation}
\begin{multline}
\omega^*\left(2+\epsiloni^*\right)
\dfrac{\mathrm{d}\vec{D}_{1,0}^*}{\mathrm{d}t^*}
\\
\shoveleft{
=
-
\left(1 - \lambdai^*-2\lambdas^*\right)
\vec{k}
-
\left(2 + \lambdai^* + 2\lambdas^*\right)
\vec{D}_{1,0}^* 
}
\\
+
\dfrac{\tau^*}{2}
\left[
 \left(1 - \epsiloni^*\right)\vec{k}
 +
 \left(2 + \epsiloni^*\right)\vec{D}_{1,0}^*
\right]
\times\left(\vec{k}\times\vec{D}_{1,0}^*\right)
,
\label{ode:D10}
\end{multline}
\begin{equation}
p_{0,0}^*
=
\dfrac{1-\epsiloni^*}{6} 
-
\dfrac{1}{3}\left(4-\epsiloni^*\right)
\vec{k}\cdot\vec{D}_{1,0}^*
-
\dfrac{1}{6}\left(2+\epsiloni^*\right)
\vec{D}_{1,0}^*\cdot\vec{D}_{1,0}^*
,
\label{p00}
\end{equation}
\begin{equation}
\hat{H}_{n,1}^*
=
0
,
\quad
\hat{B}_{n,1}^*
=
0
,
\quad
n\neq2
,
\label{eq:H20}
\end{equation}
\begin{equation}
\hat\Omega_{n,1}^*
=
0
,\quad
n\neq1
,
\end{equation}
\begin{multline}
\dfrac{\mathrm{d}\hat{H}_{2,1}^*}{\mathrm{d}t^*}
+
\dfrac{40}{19}\hat{H}_{2,1}^*
=
\left(1-\epsiloni^*\right)
\left(\vec{k}\,\vec{k}-\dfrac{1}{3}\hat{I}\right)
\\
+
\left(1+2\epsiloni^*\right)
\left[
 \left(\vec{k}\,\vec{D}_{1,0}^*\right)^\mathrm{S}
 -
 \dfrac{1}{3}\vec{k}\cdot\vec{D}_{1,0}^*\,\hat{I}
\right]
\\
+
\left(7-19\epsiloni^*\right)
\left(
 \vec{D}_{1,0}^*\,\vec{D}_{1,0}^*
 -
 \dfrac{1}{3}\vec{D}_{1,0}^*\cdot\vec{D}_{1,0}^*\,\hat{I}
\right)
,
\label{ode:H21}
\end{multline}
\begin{multline}
\hat{B}_{2,1}^*
=
\dfrac{1}{2}\left(1-\epsiloni^*\right)
\left(\vec{k}\,\vec{k}-\dfrac{1}{3}\hat{I}\right)
\\
+
\dfrac{1}{2}\left(1+2\epsiloni^*\right)
\left[
 \left(\vec{k}\,\vec{D}_{1,0}^*\right)^\mathrm{S}
 -
 \dfrac{1}{3}\vec{k}\cdot\vec{D}_{1,0}^*\,\hat{I}
\right]
\\*
-
\dfrac{1}{38}\left(11+19\epsiloni^*\right)
\left(
 \vec{D}_{1,0}^*\,\vec{D}_{1,0}^*
 -
 \dfrac{1}{3}\vec{D}_{1,0}^*\cdot\vec{D}_{1,0}^*\,\hat{I}
\right)
\\
-
\dfrac{12}{19}\hat{H}_{2,1}^*
.
\label{B21}
\end{multline}

It follows from Eq.~(\ref{ode:D10}) that
\begin{equation*}
\dfrac{\mathrm{d}\vec{D}_{1,0}^*}{\mathrm{d}t^*}
\cdot
\left(\vec{k}\times\vec{D}_{1,0}^*\right)
=
0
,
\end{equation*}
i.e., the vector $\vec{D}_{1,0}^*=\vec{D}_{1,0}^*(t^*)$, while varying, remains coplanar to the vectors $\vec{k}$ and $\vec{D}_{1,0}^*(t^*_0)$, where $t^*_0$ is some initial instant. The vector $\vec\Omega_{1,0}^*$, as it follows from Eq.~(\ref{Omega10}), remains normal to the vectors $\vec{k}$ and $\vec{D}_{1,0}^*(t^*_0)$, varying in its absolute value.

In the Cartesian basis (see Fig.~\ref{f1})
\begin{equation*}
\vec{i}_z=\vec{k}
,\quad 
\vec{i}_y
=
\dfrac
 {\vec{k}\times\vec{D}_{1,0}^*(t^*_0)}
 {\left|\vec{k}\times\vec{D}_{1,0}^*(t^*_0)\right|}
,\quad 
\vec{i}_x=\vec{i}_y\times\vec{i}_z
,
\end{equation*}
$\vec{D}_{1,0}^*={D}^*_x\vec{i}_x+{D}^*_y\vec{i}_z$, where ${D}^*_x={D}^*_x(t^*),{D}^*_z={D}^*_z(t^*)$ is the solution to the system of the ordinary differential equations
\begin{multline}
\omega^*\left(2+\epsiloni^*\right)
\dfrac{\mathrm{d}{D}_x^*}{\mathrm{d}t^*}
=
-
\left[
 2 + \lambdai^* + 2\lambdas^*
 +
 \dfrac{\tau^*}{2}\left(1 - \epsiloni^*\right)
\right]
{D}_x^*
\\
\shoveright{
-
\dfrac{\tau^*}{2}\left(2 + \epsiloni^*\right){D}_z^*{D}_x^*
,
}
\\
\shoveleft{
\omega^*\left(2+\epsiloni^*\right)
\dfrac{\mathrm{d}{D}_z^*}{\mathrm{d}t^*}
=
-
1+\lambdai^*+2\lambdas^*
-
\left(2 + \lambdai^* + 2\lambdas^*\right){D}_z^*
}
\\
+
\dfrac{\tau^*}{2}\left(2 + \epsiloni^*\right){D}_x^{*2}
.
\label{ode-D}
\end{multline}
If the following condition is satisfied
\begin{equation}
\dfrac{3}{2}
\dfrac
 {\tau^*\left(\epsiloni^*-\lambdai^*-2\lambdas^*\right)}
 {\left(2 + \lambdai^* + 2\lambdas^*\right)^2}
<
1
,
\label{knot}
\end{equation}
then the system of equations (\ref{ode-D}) has the only fixed point
\begin{equation*}
{D}_{1x}^*
=
0
,\quad
{D}_{1z}^*
=
-
\dfrac{1-\lambdai^*-2\lambdas^*}{2 + \lambdai^* + 2\lambdas^*}
, 
\end{equation*}
which is a stable knot. For the stationary solution to Eq.~(\ref{ode:D10}) corresponding to this fixed point,
\begin{equation*}
\vec{D}_{1,0}^*
=
\vec{D}_{1,0\mathrm{si}}^*
=
-
\dfrac{1-\lambdai^*-2\lambdas^*}{2 + \lambdai^* + 2\lambdas^*}
\vec{k}
,\quad
\vec\Omega_{1,0}^*
=
\vec\Omega_{1,0\mathrm{si}}^*
=
\vec0
,
\end{equation*}
i.e. the electrorotation is absent. And if
\begin{equation}
\dfrac{3}{2}
\dfrac
 {\tau^*\left(\epsiloni^*-\lambdai^*-2\lambdas^*\right)}
 {\left(2 + \lambdai^* + 2\lambdas^*\right)^2}
>
1
,
\label{focus}
\end{equation}
then the system of equations (\ref{ode-D}) has three fixed points: the saddle 
\begin{equation*}
{D}_{1x}^*
=
0
,\quad
{D}_{1z}^*
=
-
\dfrac{1-\lambdai^*-2\lambdas^*}{2 + \lambdai^* + 2\lambdas^*}
\end{equation*}
and the two stable focuses
\begin{equation*}
\begin{aligned}
{D}_{2x}^*
&=
\pm
\dfrac
 {2 + \lambdai^* + 2\lambdas^*}
 {\tau^*\left(2 + \epsiloni^*\right)}
\sqrt{
 \dfrac{3}{2}
 \dfrac
  {\tau^*\left(\epsiloni^*-\lambdai^*-2\lambdas^*\right)}
  {\left(2 + \lambdai^* + 2\lambdas^*\right)^2}
 -
 1
}
,
\\
{D}_{2z}^*
&=
-
\dfrac
 {4+2\lambdai^*+4\lambdas^*+\tau^*\left(1 - \epsiloni^*\right)}
 {\tau^*\left(2 + \epsiloni^*\right)}
\end{aligned}
\end{equation*}
symmetric with respect to the line $x=0$.
\begin{figure*}
\includegraphics{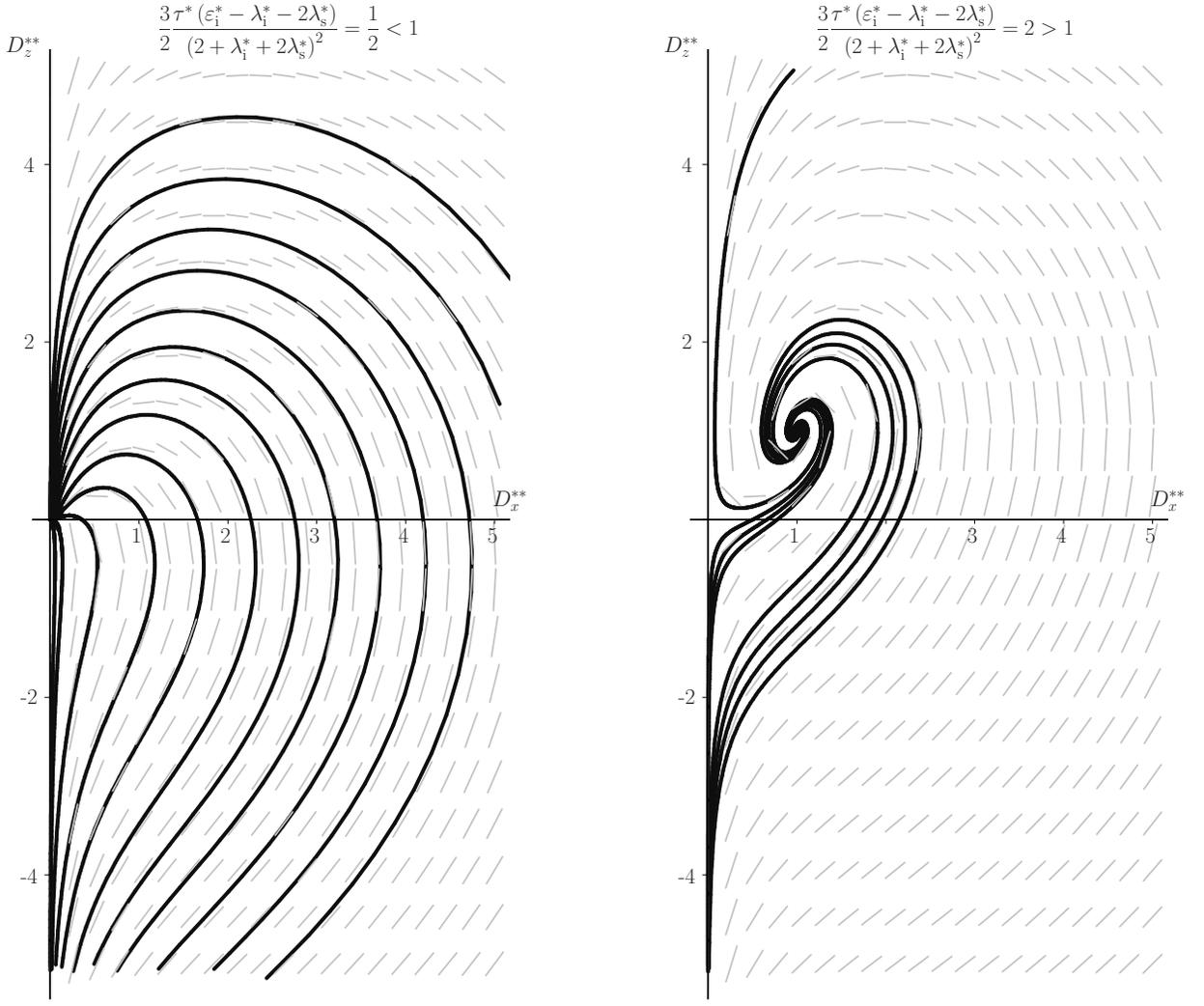}
\caption{
Phase portraits of the system of equations Eq.~(\ref{ode-D**}) for the cases without and with electrorotation.
}
\label{f2}
\end{figure*}

In Fig. \ref{f2}, the phase portraits of the system of equations
\begin{equation}
\begin{aligned}
\dfrac{\mathrm{d}{D}_x^{**}}{\mathrm{d}t^{**}}
&=
-
\left({D}_z^{**}-{D}_{0z}^{**}\right)
{D}_x^{**}
,
\\
\dfrac{\mathrm{d}{D}_z^{**}}{\mathrm{d}t^{**}}
&=
-
{D}_z^{**}
+
{D}_x^{**2}
\label{ode-D**}
\end{aligned}
\end{equation}
are depicted. The system of equations (\ref{ode-D**}) has the only parameter 
\begin{equation*}
{D}_{0z}^{**}
=
\dfrac{3\tau^*}{2}
\dfrac
 {\epsiloni^*-\lambdai^*-2\lambdas^*}
 {\left(2+\lambdai^*+2\lambdas^*\right)^2}
-
1
\end{equation*}
and is obtained from the system, Eq.~(\ref{ode-D}), with the use of the transformation
\begin{equation*}
\begin{aligned}
{D}_x^{**}
&=
\dfrac{\tau^*}{2}
\dfrac{2+\epsiloni^*}{2+\lambdai^*+2\lambdas^*}
{D}_x^*
,
\\
{D}_z^{**}
&=
\dfrac{\tau^*}{2}
\dfrac{2+\epsiloni^*}{2+\lambdai^*+2\lambdas^*}
\left(
 {D}_z^*
 +
 \dfrac{1-\lambdai^*-2\lambdas^*}{2 + \lambdai^* + 2\lambdas^*}
\right)
,
\\
{t}^{**}
&=
\dfrac
 {2+\lambdai^*+2\lambdas^*}
 {\left(2+\epsiloni^*\right)\omega^*}
t^*
\end{aligned}
\end{equation*}
that does not change the quantity and the types of the fixed points.

The 3D phase portrait is obtained from the 2D one with the rotation around $z$ axis, i.e., it has infinite number of fixed points corresponding to the stationary solutions with the electrorotation that lie on some circle. Thus, if there is no stationary solutions with the electrorotation, then, with the accuracy up to the first order with respect to $\eta^*$, all the non-stationary solutions tend to the only stationary solution without the electrorotation when the time tends to infinity. And if there exist stationary solutions with the electrorotation, the stationary solution without the electrorotation is unstable, and any non-stationary solution tends to one of the stationary solutions with the electrorotation. Note that this result is exact for the limit case $\eta^*\to0$ corresponding to a rigid sphere. 

The solution to the equation for $\hat{H}_{2,1}=\hat{H}_{2,1}(t)$, Eq.~(\ref{ode:H21}), has the form
\begin{multline}
\hat{H}_{2,1}^*(t^*)
=
\hat{H}_{2,1}^*(t^*_0)
\exp\left(-\dfrac{40}{19}t^*\right)
\\
+
\int\limits_{t^*_0}^{t^*}
 \exp\left[\dfrac{40}{19}\left({t^*}'-{t^*}\right)\right]
 \hat{F}_{2,1}^*({t^*}')
\;\mathrm{d}{t^*}'
,
\label{H21}
\end{multline}
\begin{multline}
\hat{F}_{2,1}^*(t^*)
=
\left(1-\epsiloni^*\right)
\left(\vec{k}\,\vec{k}-\dfrac{1}{3}\hat{I}\right)
\\*
+
\left(1+2\epsiloni^*\right)
\left[
 \left(\vec{k}\,\vec{D}_{1,0}^*\right)^\mathrm{S}
 -
 \dfrac{1}{3}\vec{k}\cdot\vec{D}_{1,0}^*\,\hat{I}
\right]
\\
+
\left(7-19\epsiloni^*\right)
\left(
 \vec{D}_{1,0}^*\,\vec{D}_{1,0}^*
 -
 \dfrac{1}{3}\vec{D}_{1,0}^*\cdot\vec{D}_{1,0}^*
 \,\hat{I}
\right)
,
\label{F21}
\end{multline}
where $\vec{D}_{1,0}^*=\vec{D}_{1,0}^*({t^*}')$ is the solution to Eq.~(\ref{ode:D10}). 

The equation for the surface of the drop, Eq.~(\ref{eq:S}),
can be rewritten in the form
\begin{equation*}
\dfrac
 {r^2}
 {\Rd^2
  +
  2\Rd{h}\left(\dfrac{\vec{r}}{{r}},t\right)
  +
  {h}^2\left(\dfrac{\vec{r}}{{r}},t\right)}
=
1
.
\end{equation*}
Within the small deformation approximation, i.e., when the condition, Eq.~(\ref{cond:sda}), is satisfied, this equation is equivalent to the equation
\begin{equation*}
\dfrac{r^2}{\Rd^2}
\left[
 1
 -
 2\dfrac{{h}\left(\dfrac{\vec{r}}{{r}},t\right)}{\Rd}
\right]
=
1
.
\end{equation*}
With accuracy up to the terms of the first order with respect to $\eta^*$,
\begin{equation*}
h\left(\dfrac{\vec{r}}{{r}},t\right)
=
\dfrac{1}{r^2}\hat{H}_{2,1}\stackrel{2}{\cdot}\vec{r}\,^2
.
\end{equation*}
Thus, with accuracy up to the terms of the first order over $\eta^*$, the equation for the surface of the drop has the form
\begin{equation}
\dfrac{r^2}{\Rd^2}
-
\dfrac{2\hat{H}_{2,1}}{\Rd^3}\stackrel{2}{\cdot}\vec{r}\,^2
=
1
.
\end{equation}
This is the equation of an ellipsoid with semi-axes
\begin{align}
a_1
=
\dfrac{\Rd}{\sqrt{1-2{H}_{2,1,1}\Rd^{-1}}}
\label{a1}
,\\
a_2
=
\dfrac{\Rd}{\sqrt{1-2{H}_{2,1,2}\Rd^{-1}}}
\label{a2}
,\\
a_3
=
\dfrac{\Rd}{\sqrt{1-2{H}_{2,1,3}\Rd^{-1}}}
\label{a3}
,
\end{align}
where ${H}_{2,1,1}$, ${H}_{2,1,2}$, and ${H}_{2,1,3}$ are the eigenvalues of the tensor $\hat{H}_{2,1}=\hat{H}_{2,1}(t)$.

For the stationary solutions, $\hat{H}_{2,1}^*=\hat{H}_{2,1\mathrm{s}}^*$ and $\hat{B}_{2,1}^*=\hat{B}_{2,1\mathrm{s}}^*$, where
\begin{multline}
\hat{H}_{2,1\mathrm{s}}^*
=
\dfrac{19}{40}
\left\{
\left[
 \left(1 - \epsiloni^*\right)
 \vec{k}
 +
 \left(2 + \epsiloni^*\right) \vec{D}_{1,0\mathrm{s}}^*
\right]
\,
\left(\vec{k} - \vec{D}_{1,0\mathrm{s}}^*\right)
\right\}^\mathrm{S}
\\*
-
\dfrac{19}{120}
\left[
 \left(1 - \epsiloni^*\right)
 \vec{k}
 +
 \left(2 + \epsiloni^*\right) \vec{D}_{1,0\mathrm{s}}^*
\right]
\cdot
\left(\vec{k} - \vec{D}_{1,0\mathrm{s}}^*\right)
\;\hat{I}
\\*
+
\dfrac{9}{8}
\left(
 \vec{D}_{1,0\mathrm{s}}^*\,\vec{D}_{1,0\mathrm{s}}^*
 -
 \dfrac{1}{3}
 \vec{D}_{1,0\mathrm{s}}^*\cdot\vec{D}_{1,0\mathrm{s}}^*\,
 \hat{I}
\right)
,
\label{H21s}
\end{multline}
\begin{multline}
\hat{B}_{2,1\mathrm{s}}^*
=
\dfrac{1}{5}
\left\{
 \left[
  \left(1 - \epsiloni^*\right)
  \vec{k}
  +
  \left(2 + \epsiloni^*\right) \vec{D}_{1,0\mathrm{s}}^*
 \right]
 \;
 \left(\vec{k} - \vec{D}_{1,0\mathrm{s}}^*\right)
\right\}^\mathrm{S}
\\
-
\dfrac{1}{15}
\left[
 \left(1 - \epsiloni^*\right)
 \vec{k}
 +
 \left(2 + \epsiloni^*\right) \vec{D}_{1,0\mathrm{s}}^*
\right]
\cdot
\left(\vec{k} - \vec{D}_{1,0\mathrm{s}}^*\right)
\hat{I}
,
\end{multline}
and $\vec{D}_{1,0\mathrm{s}}^*$ is one of the stationary solutions to the equation for $\vec{D}_{1,0}^*(t^*)$, Eq.~(\ref{ode:D10}). For the stationary solutions with the electrorotation, $\vec\Omega_{1,0}^*=\vec\Omega_{1,0\mathrm{sr}}^*$, 
\begin{multline}
\Omega_{1,0\mathrm{sr}}^*
=
\dfrac
 {2 + \lambdai^* + 2\lambdas^*}
 {\tau^*\left(2 + \epsiloni^*\right)}
\sqrt{
 \dfrac{3}{2}
 \dfrac
  {\tau^*\left(\epsiloni^* - \lambdai^* - 2\lambdas^*\right)}
  {\left(2 + \lambdai^* + 2\lambdas^*\right)^2}
 -
 1
}
,
\\
\vec\Omega_{1,0\mathrm{sr}}^*\cdot\vec{k}
=
0
,
\end{multline}
\begin{multline}
\vec{D}_{1,0}^*
=
\vec{D}_{1,0\mathrm{sr}}^*
=
-
\left(
 \dfrac{2}{\tau^*}
 \dfrac{2 + \lambdai^* + 2\lambdas^*}{2 + \epsiloni^*}
 +
 \dfrac{1 - \epsiloni^*}{2 + \epsiloni^*}
\right)
\vec{k}
\\
+
2\vec{k}\times\vec\Omega_{1,0\mathrm{sr}}^*
,
\end{multline}
\begin{equation}
{H}_{2,1,1}^*
=
{H}_{2\mathrm{sr}}^*
=
\dfrac{1}{6}\left({a}_H+{c}_H\right)
-
\dfrac{1}{2}\sqrt{\left({c}_H-{a}_H\right)^2+4{b}_H^2}
,
\label{H211}
\end{equation}
\begin{equation}
{H}_{2,1,2}^*
=
{H}_{3\mathrm{sr}}^*
=
-
\dfrac{1}{3}\left({a}_H+{c}_H\right)
,
\label{H212}
\end{equation}
\begin{equation}
{H}_{2,1,3}^*
=
{H}_{1\mathrm{sr}}^*
=
\dfrac{1}{6}\left({a}_H+{c}_H\right)
+
\dfrac{1}{2}\sqrt{\left({c}_H-{a}_H\right)^2+4{b}_H^2}
,
\label{H213}
\end{equation}
\begin{equation}
{a}_H
=
\dfrac{1}{40}\left(7-19\epsiloni^*\right)
\left(\vec{D}_{1,0\mathrm{sr}}^*\cdot\vec{i}_x\right)^2
,
\label{aH}
\end{equation}
\begin{multline}
{b}_H
=
\dfrac{19}{40}\left(1+2\epsiloni^*\right)
\vec{D}_{1,0\mathrm{sr}}^*\cdot\vec{i}_x
\\
+
\dfrac{1}{20}\left(7-19\epsiloni^*\right)
\vec{D}_{1,0\mathrm{sr}}^*\cdot\vec{k}\;
\vec{D}_{1,0\mathrm{sr}}^*\cdot\vec{i}_x
,
\label{bH}
\end{multline}
\begin{multline}
{c}_H
=
\dfrac{19}{40}\left(1-\epsiloni^*\right)
+
\dfrac{19}{40}\left(1+2\epsiloni^*\right)
\vec{D}_{1,0\mathrm{sr}}^*\cdot\vec{k}
\\
+
\dfrac{1}{40}\left(7-19\epsiloni^*\right)
\left(\vec{D}_{1,0\mathrm{sr}}^*\cdot\vec{k}\right)^2
.
\label{cH}
\end{multline}

In the dimensional form, the condition, Eq.(\ref{focus}), can be written in the form
\begin{equation}
\Ea>{E}_\mathrm{c}
,
\label{c:ER}
\end{equation}
where ${E}_\mathrm{c}$ is the critical value of the intensity defined as follows:
\begin{equation}
{E}_\mathrm{c}
=
\dfrac
 {2
 +
 \dfrac{\lambdai}{\lambdae}
 +
 \dfrac{2\lambdas}{\Rd\lambdae}}
 {\sqrt{
   \dfrac{\epsiloni}{\epsilone}
   -
   \dfrac{\lambdai}{\lambdae}
   -
   \dfrac{2\lambdas}{\Rd\lambdae}
 }}
\sqrt{\dfrac{32\pi^2\lambdae\etae}{3\epsilone^2}}
.
\label{Ec}
\end{equation}
Note that ${E}_\mathrm{c}$ is determined only if the following condition is fulfilled:
\begin{equation}
\dfrac{\epsiloni}{\epsilone}
>
\dfrac{\lambdai}{\lambdae}
+
\dfrac{2\lambdas}{\Rd\lambdae}
.
\label{c:ER-nes}
\end{equation}

\begin{figure}
\includegraphics{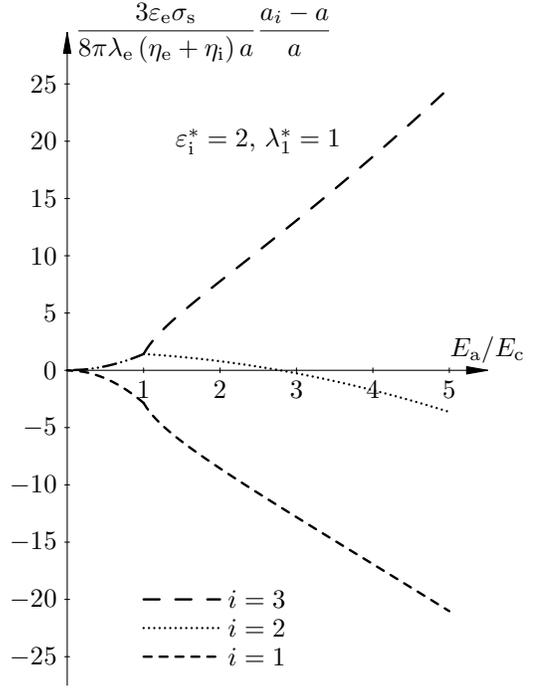}
\caption{
Dependencies of the semi-axes on the intensity of the applied electric field
}
\label{f3}
\end{figure}

\begin{figure}
\includegraphics{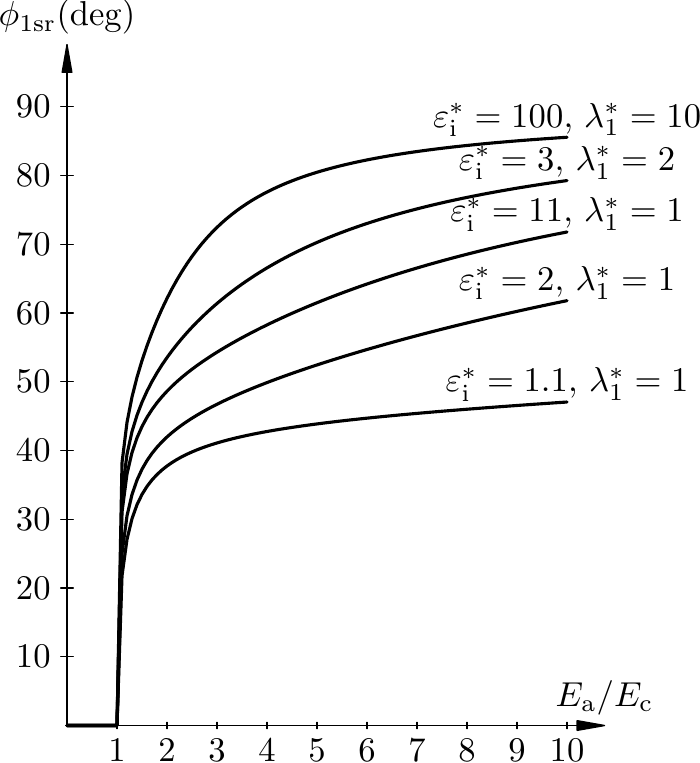}
\caption{
Angle between the minor semi-axis and the intensity vector of the applied electric field
}
\label{f4}
\end{figure}

For sufficiently small deformations of the drop, the dependencies of the of the minor, medium, and major semi-axes, ${a}_1$, ${a}_2$, and ${a}_3$, on the dimensionless intensity of the applied electric field, $\Ea^*=\Ea/{E}_\mathrm{c}$, can be written in the form
\begin{equation}
\dfrac{a_i-\Rd}{\Rd}
=
\dfrac
 {8\pi\lambdae\left(\etae+\etai\right)\Rd}
 {3\epsilone\sigma_\mathrm{s}}
\dfrac{\left(2+\lambda_1^*\right)^2}{\epsiloni^*-\lambda_1^*}
\Ea^{*2}{H}_{2,1,i}^*
,\ 
i=1,2,3
,
\end{equation}
where
\begin{equation}
\lambda_1^*
=
\lambdai^*
+
2\lambdas^*
\label{lambda1}
\end{equation}
and the replacement
\begin{equation}
\tau^*
=
\dfrac{2}{3}
\dfrac
 {\left(2+\lambda_1^*\right)^2\Ea^{*2}}
 {\epsiloni^*-\lambda_1^*}
\end{equation}
is done in the expressions for ${H}_{2,1,i}^*$. The medium semi-axis is directed along the $y$ axis. The minor and major semi-axes are in the $xz$ plane and make the angles
\newcommand{\arccot}{\mathop\mathrm{arccot}}
\begin{equation}
\phi_{1\mathrm{sr}}
=
\dfrac{1}{2}
\arccot\dfrac{{c}_H-{a}_H}{2{b}_H}
,\quad
\phi_{3\mathrm{sr}}
=
\dfrac{\pi}{2}
-
\phi_{1\mathrm{sr}}
\label{phi1,phi3}
\end{equation}
with $\vec{k}$. If $\epsiloni^*\neq1$, the ellipsoid tends to take the shape of a prolate spheroid as the intensity of the applied electric field tends to infinity. The dependencies of ${a}_1$, ${a}_2$, ${a}_3$, and $\phi_{1\mathrm{sr}}$ on the intensity for drops that are oblate spheroids below the critical value are depicted in Figs. \ref{f3} and \ref{f4}.

\subsection{Corrections of the first and second orders}

Using the boundary conditions (\ref{BC:j}), (\ref{bc:sn}), and (\ref{bc:st}) in the dimensionless form and Eqs.~(\ref{Omegan0,Dn0})--(\ref{B21}), one obtains the expressions and relations for $\hat{D}_{n,1}^*$, $\vec\Omega_{1,1}^*$, $p_{0,1}^*$, $\hat\Omega_{n,2}^*$ ($n\neq1$), $\hat{H}_{n,2}^*$, and $\hat{B}_{n,2}^*$, Eqs. (\ref{Omega11})--(\ref{Omega32}), which are written down in \ref{A2}.

The linear ordinary differential equations for $\hat{H}_{2,2}^*$ and $\hat{H}_{4,2}^*$, Eqs.~(\ref{ode:H22}) and  (\ref{ode:H42}), have solutions (\ref{H22})--(\ref{F42}), from which it follows that $\vec\Omega_{1,1}^*$, $\hat{H}_{2,2}^*$, $\hat{H}_{4,2}^*$, $\hat{B}_{2,2}^*$, $\hat{B}_{4,2}^*$, and $\hat\Omega_{3,2}^*$ are determined by $\vec{D}_{1,1}^*(t^*)$, $\hat{D}_{3,1}^*(t^*)$, and $\vec{D}_{1,0}^*(t^*)$, which are the solutions to the differential equations (\ref{ode:D11}), (\ref{ode:D31}), and (\ref{ode:D10}).

As $t^*\to\infty$, the solutions of Eqs.~(\ref{ode:D11}) and (\ref{ode:D31}) tend to the stationary solutions $\vec{D}_{1,1}^*(t^*)=\vec{D}_{1,1\mathrm{sr}}^*$ and $\hat{D}_{3,1}^*(t^*)=\hat{D}_{3,1\mathrm{sr}}^*$, Eqs.~(\ref{D11sr})---(\ref{G31sr}). Thus, these stationary solutions are always stable.

\begin{figure}
\includegraphics{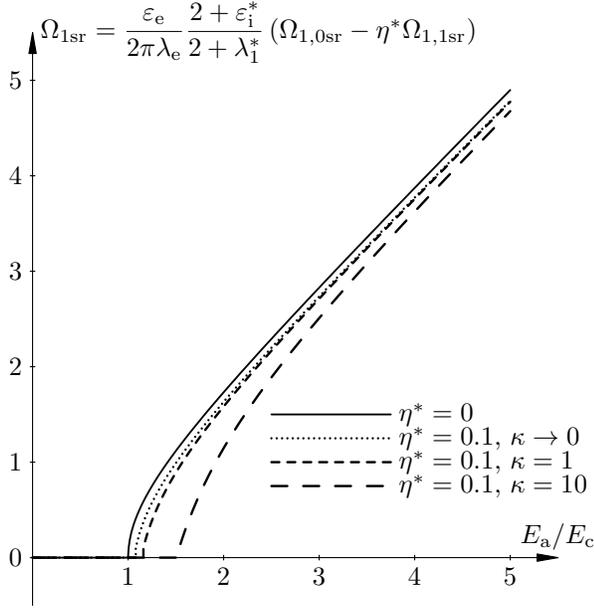}
\caption{
$\Omega_{1\mathrm{sr}}$ with the accuracy up to the term of the first order with respect to $\eta^*$ at various values of $\kappa=(\epsiloni^*-\lambda_1^*)/(\lambda_1^*+2)$
}
\label{f5}
\end{figure}
\begin{figure}
\includegraphics{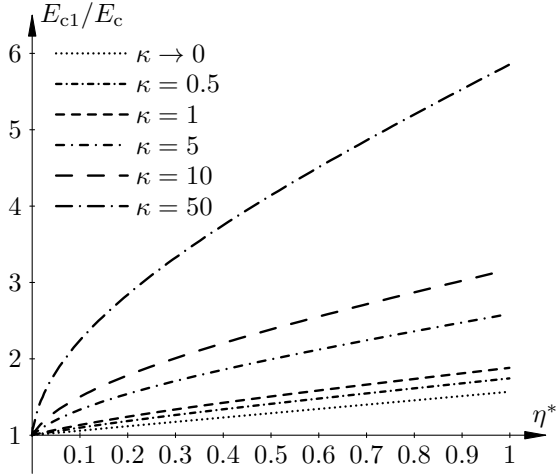}
\caption{Dependence of $E_{\mathrm{c}1}^*$ on $\eta^*$ at various values of $\kappa=(\epsiloni^*-\lambda_1^*)/(\lambda_1^*+2)$}
\label{f6}
\end{figure}
Introducing a new dimensionless angular velocity of the electrorotation in order to study its dependence on $\Ea^*=\Ea/E_\mathrm{c}$,
\begin{equation}
\vec\Omega_1
=
\dfrac{2\pi\lambdae}{\epsilone}
\dfrac{2+\lambda_1^*}{2+\epsiloni^*}
\vec\Omega_1^{**}
,
\end{equation}
one obtains
\begin{equation}
\vec\Omega_{1,0\mathrm{sr}}^{**}
=
-
\sqrt{\Ea^{*2}-1}\;\vec{i}_y
,
\end{equation}
\begin{equation}
\vec\Omega_{1,1\mathrm{sr}}^{**}
=
\dfrac
 {3
  \left[
   \left(3\Ea^{*2}+17\right)\Ea^{*2}
   +
   8
   \dfrac{\epsiloni^*-\lambda_1^*}{2+\lambda_1^*}
   \left(\Ea^{*2}+3\right)
  \right]}
 {50\Ea^{*2}\sqrt{\Ea^{*2}-1}}
\vec{i}_y
.
\end{equation}
The dependence of $\Omega_{1\mathrm{sr}}$ found with accuracy up to the term of the first order with respect to $\eta^*$ on the intensity of the applied electric field is shown in Fig. \ref{f5}. Note that the correction of the first order to the angular velocity tends to infinity as $\Ea^*\to1$, i.e., it is not valid near the critical value of the intensity. This fact may be interpreted as pointing to the increase of the critical value of the intensity when $\eta^*$ increases from zero to some finite value. It may be assumed that the critical value of $E_\mathrm{a}$ in the first order approximation with respect to $\eta^*$, $E_{\mathrm{c}1}$, is determined by the equation
\begin{equation}
\Omega_{1,0\mathrm{sr}}(E_{\mathrm{c}1}^*)
-
\eta^*\Omega_{1,1\mathrm{sr}}(E_{\mathrm{c}1}^*)
=
0
,
\label{Ec1}
\end{equation}
where $E_{\mathrm{c}1}^*=E_{\mathrm{c}1}/E_{\mathrm{c}}$. The dependence of $E_{\mathrm{c}1}^*$ determined by Eq.~(\ref{Ec1}) on $\eta^*$ is shown in Fig.~\ref{f6}.

\section{Summary and discussion}

Electric field intensity and velocity and pressure in the electrohydrodynamic flow inside and outside a drop of a viscous weakly conducting polarizable liquid suspended in another viscous weakly conducting polarizable liquid immiscible with the former for the case of electrorotation are found in the small deformation approximation in the form of the series, Eqs.(\ref{E})--(\ref{phi}) and (\ref{v,p})--(\ref{pi}), determined by the functions $\hat{D}_n(t)$, $\hat\Omega_n(t)$, $\hat{B}_n(t)$ and $p_0(t)$. The variation of the shape of the drop is determined by the functions $\hat{H}_n(t)$ according to Eqs.(\ref{eq:S}) and (\ref{h}). The functions $\hat{D}_n(t)$, $\hat\Omega_n(t)$, $\hat{B}_n(t)$, $\hat{H}_n(t)$, and $p_0(t)$ adimensionalized according to Eq.(\ref{adim}) can be presented in the form of the asymptotic expansions, Eq.(\ref{asym}), in the parameter $\eta^*$. The relations, Eqs.(\ref{Omegan0,Dn0})--(\ref{B21}) and (\ref{Omega11})--(\ref{Omega32}), are obtained for the terms of the zeroth, first, and second orders of these expansions.

It is established that the electric field in the zeroth approximation is determined by the dipole moment of the drop and the correction of the first order the the field is determined by  the correction of the first order to the dipole moment and by the octupole moment. Up to the terms of the first order, the electric dipole and octupole moments of the drop are determined by the differential equations (\ref{ode:D10}), (\ref{ode:D11}), and (\ref{ode:D31}) and the angular velocity of electrorotation is determined by Eqs.(\ref{Omega10}) and (\ref{Omega11}). The shape of the drop is determined up to the terms of the second order by the differential equations (\ref{ode:H21}), (\ref{ode:H22}), and (\ref{ode:H42}). The stationary solutions of the differential equations mentioned above are found in the explicit form and their stability is investigated.

It is established that, in the zeroth approximation, besides the stationary solution without electrorotation, which always exists, there also exist stationary solutions with electrorotation determined with the accuracy up to the direction of the angular velocity of electrorotation when the condition, Eq.(\ref{focus}), is fulfilled. All the solutions with electrorotation are stable and the solution without electrorotation are unstable if $\Ea>{E}_\mathrm{c}$. If $\Ea<{E}_\mathrm{c}$, then the solution without electrorotation is stable.

It is established that, up to the terms of the first order, the stationary shape of the drop is an ellipsoid of the general form the medium axis of which is directed along the angular velocity and major and minor ones make some angles with the electric field intensity vector. The explicit expressions are obtained for the semiaxes of the ellipsoid, Eqs.(\ref{aH})--(\ref{cH}), and for these angles, Eq.(\ref{phi1,phi3}). Up to the terms of the second order, the surface of the drop is described by an equation of the fourth order.

The influence of the surface conductivity on the leading terms of the asymptotic expansions can be taken into account by the replacement $\lambdai^*\to\lambda_1^*=\lambdai^*+2\lambdas^*$, but this rule is not always applicable for the terms of higher order [see Eqs.(\ref{ode:D31}) and (\ref{D31sr})].

Since a stable stationary solution always exists in the zeroth approximation and the corrections to it of the first and second orders are also stable, one may assume that no transition to non-decaying unsteady solutions takes place when the conditions for the approximations used in solving the problem are fulfilled. Thus, in order to better understand the experiments where such a transition, described by the theory of the work \cite{Salipante&Vlahovska13}, was observed, it is necessary to refuse to use at least one of the approximations: the quasistationary field approximation, quasisteady flow approximation, or small deformation approximation. The frequency of the oscillations observed in the experiments was too small in order the conditions for the quasistationary field and quasisteady flow approximations not to be fulfilled. This fact provides the basis to assume that the transition to the deformational oscillations takes place when the conditions for the small deformation approximation are not fulfilled. Thus, one may hope that a solution with the transition to the deformational oscillations can be obtained if the influence of the deviation of the shape of the drop from a sphere on the electric field and on the electrohydrodynamic flow is taken into account as it was done in Ref. \onlinecite{Ajayi}.

\acknowledgments

Partially supported by Russian Foundation for Basic Research grants Nos. 16-01-00157 and 17-01-00037.

\appendix
\section{Lamb' solution}
\label{A1}

The solution to the continuity equation (\ref{de}) and Navier--Stokes equations in the low Reynolds number approximation, Eq.~(\ref{me}), can be constructed from the general Lamb's solution (cf. Ref. \onlinecite{Lamb}, art. 336) for a quasi-steady flow
\begin{equation}
\vec{v}
=
\vec{v}_1
+
\vec{v}_2
+
\vec{v}_3
,
\end{equation}
\begin{multline}
\vec{v}_1
=
\sum\limits_{n=1}^\infty
\nabla \times
\left\{
 \left[
  \hat{A}_{1,1,n}
  \stackrel{n-1}{\cdot} \dfrac{\vec{r}\,^{n-1}}{\Rd^{n-1}}
 \right]
 \times\vec{r}
\right\}
\\
+
\sum\limits_{n=1}^\infty
\nabla \times
\left\{
 \left[
  \hat{A}_{1,2,n}
  \stackrel{n-1}{\cdot} \dfrac{\vec{r}\,^{n-1}\Rd^{n+2}}{r^{2n+1}}
 \right]
 \times\vec{r}
\right\}
,
\end{multline}
\begin{multline}
\vec{v}_2
=
\sum\limits_{n=1}^\infty
\nabla \times
\left\{
 \left[
  \hat{A}_{2,1,n}
  \stackrel{n-1}{\cdot} \dfrac{\vec{r}\,^{n-1} r^2}{\Rd^{n+1}}
 \right]
 \times\vec{r}
\right\}
\\
+
\sum\limits_{n=1}^\infty
\nabla \times
\left\{
 \left[
  \hat{A}_{2,2,n}
  \stackrel{n-1}{\cdot} \dfrac{\vec{r}\,^{n-1}\Rd^n}{r^{2n-1}}
 \right]
 \times\vec{r}
\right\}
,
\end{multline}
\begin{multline}
\vec{v}_3
=
\sum\limits_{n=1}^\infty
 n
 \left[
  \hat{A}_{3,1,n}
  \stackrel{n-1}{\cdot} \dfrac{\vec{r}\,^{n-1}}{\Rd^{n}}
 \right]
 \times\vec{r}
\\
+
\sum\limits_{n=1}^\infty
 n
 \left[
  \hat{A}_{3,2,n}
  \stackrel{n-1}{\cdot} \dfrac{\vec{r}\,^{n-1}\Rd^{n+1}}{r^{2n+1}}
 \right]
 \times\vec{r}
,
\end{multline}
\begin{multline}
p
=
\eta
\sum\limits_{n=1}^\infty
 \dfrac{(2n+2)(2n+3)}{n} \hat{A}_{2,1,n} \stackrel{n}{\cdot}
 \dfrac{\vec{r}\,^n}{\Rd^{n+1}}
\\
+
\eta
\sum\limits_{n=1}^\infty
 2(2n-1) \hat{A}_{2,2,n} \stackrel{n}{\cdot}
 \dfrac{\vec{r}\,^n\Rd^n}{r^{2n+1}}
.
\end{multline}
Here, $\hat{A}_{i,j,n}=\hat{A}_{i,j,n}(t)$ ($i=1,2,3$, $j=1,2$, $n=1,2,\dots$) are some tensors depending on the time each of which is either an arbitrary vector for $n=1$ or an arbitrary irreducible tensor for $n>1$.
This solution satisfies the continuity equation (\ref{de}) and the Navier--Stokes equation in the low Reynolds number approximation, Eq.~(\ref{me}). Determining differently the tensor coefficients $\hat{A}_{i,j,n}$ inside ($r<\Rd$) and outside ($r>\Rd$) the drop and satisfying the  boundedness conditions, boundary conditions at infinity, Eqs.~(\ref{BC:inf:v})--(\ref{BC:inf:p}), and impenetrability, Eq.~(\ref{BC:vn}), and no-slip, Eq.~(\ref{BC:vt}), conditions, one obtains
\begin{equation}
\vec{v}
=
\begin{cases}
\vec{v}_\mathrm{e}
,\quad
r>\Rd
,
\\
\vec{v}_\mathrm{s}
,\quad
r=\Rd
,
\\
\vec{v}_\mathrm{i}
,\quad
r<\Rd
,
\end{cases}
\quad
p
=
\begin{cases}
{p}_\mathrm{e}
,\quad
r>\Rd
,
\\
{p}_\mathrm{i}
,\quad
r<\Rd
,
\end{cases}
\label{v,p}
\end{equation}
\begin{multline}
\vec{v}_\mathrm{e}
=
\sum\limits_{n=1}^\infty
 (n+1)\left(r^2-\Rd^2\right)\Rd^n
 \hat{B}_n \stackrel{n}{\cdot} \dfrac{\vec{r}\,^{n+1}}{r^{2n+3}}
\\
+
\sum\limits_{n=1}^\infty
 (n+1)\left[n r^2-(n-2)\Rd^2\right]\Rd^n
 \hat{C}_n \stackrel{n}{\cdot} \dfrac{\vec{r}\,^{n+1}}{r^{2n+3}}
\\
+
\sum\limits_{n=1}^\infty
 \dfrac{\vec{r}}{r}\times
 \left(
  \left\{
   \left[(-n+2)r^2 + n\Rd^2\right]\Rd^n
   \hat{B}_n \stackrel{n-1}{\cdot} \dfrac{\vec{r}\,^{n-1}}{r^{2n+1}}
  \right\}
  \times\dfrac{\vec{r}}{r}
 \right)
\\
-
\sum\limits_{n=1}^\infty
 \dfrac{\vec{r}}{r}\times
 \left(
  \left\{
   (n-2)n\left(r^2-\Rd^2\right)\Rd^n
   \hat{C}_n \stackrel{n-1}{\cdot} \dfrac{\vec{r}\,^{n-1}}{r^{2n+1}}
  \right\}
  \times\dfrac{\vec{r}}{r}
 \right)
\\
+
\sum\limits_{n=1}^\infty
 n
 \left[
  \hat\Omega_n
  \stackrel{n-1}{\cdot} \dfrac{\vec{r}\,^{n-1}\Rd^{n+2}}{r^{2n+1}}
 \right]
 \times\vec{r}
,
\label{ve}
\end{multline}
\begin{multline}
\vec{v}_\mathrm{i}
=
\sum\limits_{n=1}^\infty
 (n+1)\left(r^2-\Rd^2\right)\dfrac{1}{r^2}
 \hat{B}_n \stackrel{n}{\cdot} \dfrac{\vec{r}\,^{n+1}}{\Rd^{n+1}}
\\
-
\sum\limits_{n=1}^\infty
 (n+1)\left[(n+1)r^2-(n+3)\Rd^2\right]\dfrac{1}{r^2}
 \hat{C}_n \stackrel{n}{\cdot} \dfrac{\vec{r}\,^{n+1}}{\Rd^{n+1}}
\\
+
\sum\limits_{n=1}^\infty
 \dfrac{\vec{r}}{r}\times
 \left(
  \left\{
   \left[(n+3)r^2 - (n+1)\Rd^2\right]
   \hat{B}_n \stackrel{n-1}{\cdot} \dfrac{\vec{r}\,^{n-1}}{\Rd^{n+1}}
  \right\}
  \times\dfrac{\vec{r}}{r}
 \right)
\\
-
\sum\limits_{n=1}^\infty
 \dfrac{\vec{r}}{r}\times
 \left(
  \left\{
   (n+1)(n+3)\left(r^2-\Rd^2\right)
   \hat{C}_n \stackrel{n-1}{\cdot} \dfrac{\vec{r}\,^{n-1}}{\Rd^{n+1}}
  \right\}
  \times\dfrac{\vec{r}}{r}
 \right)
\\
+
\sum\limits_{n=1}^\infty
 n
 \left[
  \hat\Omega_n
  \stackrel{n-1}{\cdot} \dfrac{\vec{r}\,^{n-1}}{\Rd^{n-1}}
 \right]
 \times\vec{r}
,
\end{multline}
\begin{multline}
\vec{v}_\mathrm{s}
=
\sum\limits_{n=1}^\infty
 2(n+1) \hat{C}_n \stackrel{n}{\cdot} \vec{n}\,^{n+1}
\\
+
\sum\limits_{n=1}^\infty
 2
 \vec{n}\times
 \left\{
  \left[\hat{B}_n \stackrel{n-1}{\cdot} \vec{n}\,^{n-1}\right]
  \times\vec{n}
 \right\}
\\
+
\sum\limits_{n=1}^\infty
 n
 \left[
  \hat\Omega_n
  \stackrel{n-1}{\cdot} \vec{n}\,^{n-1}
 \right]
 \times\vec{n}
,
\label{vs}
\end{multline}
\begin{equation}
p_\mathrm{e}
=
\etae 
\sum\limits_{n=1}^\infty
 2(2n-1) \left(\hat{B}_n+n\hat{C}_n\right) \stackrel{n}{\cdot}
 \dfrac{\vec{r}\,^n\Rd^n}{r^{2n+1}}
+
p_\infty
,
\end{equation}
\begin{multline}
p_\mathrm{i}
=
\etai 
\sum\limits_{n=1}^\infty
 \dfrac{(2n+2)(2n+3)}{n} 
 \left[\hat{B}_n-(n+1)\hat{C}_n\right] \stackrel{n}{\cdot}
 \dfrac{\vec{r}\,^n}{\Rd^{n+1}}
\\
+
p_0
.
\label{pi}
\end{multline}
Here, $\hat\Omega_{n}=\hat\Omega_{n}(t)$, $\hat{B}_{n}=\hat{B}_{n}(t)$, $\hat{C}_{n}=\hat{C}_{n}(t)$ are some tensors depending on the time each of which is either an arbitrary vector for $n=1$ or an arbitrary irreducible tensor for $n>1$, $p_0=p_0(t)$ is the pressure at the center of the drop, which is to be found. 

\section{Corrections of the first and second orders}
\label{A2}

Using the boundary conditions Eqs.~(\ref{BC:j}), (\ref{bc:sn}), and (\ref{bc:st}) in the dimensionless form and the relations Eqs.~(\ref{Omegan0,Dn0})--(\ref{B21}), one obtains the following expressions and relations for $\hat{D}_{n,1}^*$, $\vec\Omega_{1,1}^*$, $p_{0,1}^*$, $\hat\Omega_{n,2}^*$ ($n\neq1$), $\hat{H}_{n,2}^*$, and $\hat{B}_{n,2}^*$:
\begin{equation}
\vec\Omega_{1,1}^*
=
-
\dfrac{1}{2}\vec{k}\times\vec{D}_{1,1}^*
,
\label{Omega11}
\end{equation}
\begin{equation}
\hat{D}_{n,1}^*
=
0
,
\quad
n\neq1
,\ 
n\neq3
,
\label{Dn1}
\end{equation}
\begin{multline}
\omega^*\left(2+\epsiloni^*\right)
\dfrac{\mathrm{d}\vec{D}_{1,1}^*}{\mathrm{d}t^*}
+
\left(
 2
 +
 \lambdai^*
 +
 2\lambdas^*
\right)
\vec{D}_{1,1}^*
\\*
-
\dfrac{\tau^*}{2}
\left[
 \left(1 - \epsiloni^*\right)\vec{k}
 +
 \left(2 + \epsiloni^*\right)
 \vec{D}_{1,0}^*
\right]
\times\left(\vec{k}\times\vec{D}_{1,1}^*\right)
\\
-
\tau^*
\left(2+\epsiloni^*\right)
\vec\Omega_{1,0}^*\times\vec{D}_{1,1}^*
\\*
\shoveleft{
=
\dfrac{3}{5}
\tau^*
\left[
 \left(1-\epsiloni^*\right)\vec{k}
 +
 \left(2+\epsiloni^*\right)\vec{D}_{1,0}^*
\right]
\cdot\hat{B}_{2,1}^*
}
\\*
-
\dfrac{2}{5}
\tau^*
\left[
 \left(1-\epsiloni^*\right)\vec{k}
 +
 \left(2+\epsiloni^*\right)\vec{D}_{1,0}^*
\right]
\cdot\dfrac{\mathrm{d}\hat{H}_{2,1}^*}{\mathrm{d}t^*}
\label{ode:D11}
,
\end{multline}
\begin{multline}
\omega^*\left(4+3\epsiloni^*\right)
\dfrac{\mathrm{d}\hat{D}_{3,1}^*}{\mathrm{d}t^*}
+
\left(4+3\lambdai^*+12\lambdas^*\right)
\hat{D}_{3,1}^*
\\
-
3\tau^*
\left(4+3\epsiloni^*\right)
\left(\vec\Omega_{1,0}^*\times\hat{D}_{3,1}^*\right)^\mathrm{S}
\\*
\shoveleft{
=
-
\tau^*
\left(
\left\{
 \left[
  \left(1-\epsiloni^*\right)\vec{k}
  +
  \left(2+\epsiloni^*\right)\vec{D}_{1,0}^*
 \right]
 \;\dfrac{\mathrm{d}\hat{H}_{2,1}^*}{\mathrm{d}t^*}
\right\}^\mathrm{S}
\right.
}
\\*
\left.
-
\dfrac{2}{5}
\left\{
 \left[
  \left(1-\epsiloni^*\right)\vec{k}
  +
  \left(2+\epsiloni^*\right)\vec{D}_{1,0}^*
 \right]
 \cdot\dfrac{\mathrm{d}\hat{H}_{2,1}^*}{\mathrm{d}t^*}\;\hat{I}
\right\}^\mathrm{S}
\right)
\\*
+
4\tau^*
\left(
\left\{
 \left[
  \left(1-\epsiloni^*\right)\vec{k}
  +
  \left(2+\epsiloni^*\right)\vec{D}_{1,0}^*
 \right]
 \;\hat{B}_{2,1}^*
\right\}^\mathrm{S}
\right.
\\*
\left.
-
\dfrac{2}{5}
\left\{
 \left[
  \left(1-\epsiloni^*\right)\vec{k}
  +
  \left(2+\epsiloni^*\right)\vec{D}_{1,0}^*
 \right]
 \cdot
 \hat{B}_{2,1}^*
 \;\hat{I}
\right\}^\mathrm{S}
\right)
,
\label{ode:D31}
\end{multline}
\begin{equation}
p_{0,1}^*
=
-
\dfrac{1}{3}
\left[
 \left(4-\epsiloni^*\right)\vec{k}
 +
 \left(2-\epsiloni^*\right)\vec{D}_{1,0}^*
\right]
\cdot\vec{D}_{1,1}^*
\label{p01}
,
\end{equation}
\begin{equation}
\hat{H}_{n,2}^*
=
0
,
\quad
\hat{B}_{n,2}^*
=
0
,
\quad
n\neq2
,\ 
n\neq4
,
\label{Hn2,Bn2}
\end{equation}
\begin{equation}
\hat\Omega_{n,2}^*
=
0
,
\quad
n\neq1
,\ 
n\neq3
,
\label{Omegan2}
\end{equation}
\begin{multline}
\dfrac{\mathrm{d}\hat{H}_{2,2}^*}{\mathrm{d}t^*}
+
\dfrac{40}{19}\hat{H}_{2,2}^*
=
-
\dfrac{15}{19}\hat{B}_{2,1}^*
+
\dfrac{68}{57}\dfrac{\mathrm{d}\hat{H}_{2,1}^*}{\mathrm{d}t^*}
\\*
+
\dfrac{1}{19}
\left\{
 \left[
  19\left(1+2\epsiloni^*\right)\vec{k}
  -
  2\left(7+19\epsiloni^*\right)\vec{D}_{1,0}^*
 \right]
 \vec{D}_{1,1}^*
\right\}^\mathrm{S}
\\*
-
\dfrac{1}{57}
 \left[
  19\left(1+2\epsiloni^*\right)\vec{k}
  -
  2\left(7+19\epsiloni^*\right)\vec{D}_{1,0}^*
 \right]
\cdot\vec{D}_{1,1}^*\;\hat{I}
\\*
+
\dfrac{12}{133}
\left[
 \left(8-\epsiloni^*\right)\vec{k}
 +
 \left(39-34\epsiloni^*\right)\vec{D}_{1,0}^*
\right]
\cdot\hat{D}_{3,1}^*
,
\label{ode:H22}
\end{multline}
\begin{multline}
\dfrac{\mathrm{d}\hat{H}_{4,2}^*}{\mathrm{d}t^*}
+
\dfrac{72}{17}\hat{H}_{4,2}^*
\\
=
\dfrac{1}{51}
\left\{
 \left[
  17\left(1+6\epsiloni^*\right)\vec{k}
  +
  2\left(41-51\epsiloni^*\right)\vec{D}_{1,0}^*
 \right]
 \;\hat{D}_{3,1}^*
\right\}^\mathrm{S}
\\*
-
\dfrac{1}{119}
\left\{
 \left[
  17\left(1+6\epsiloni^*\right)\vec{k}
  +
  2\left(41-51\epsiloni^*\right)\vec{D}_{1,0}^*
 \right]
 \cdot\hat{D}_{3,1}^*\;\hat{I}
\right\}^\mathrm{S}
,
\label{ode:H42}
\end{multline}
\begin{multline}
\hat{B}_{2,2}^*
=
-
\dfrac{9}{38}\hat{B}_{2,1}^*
+
\dfrac{25}{38}\dfrac{\mathrm{d}\hat{H}_{2,1}^*}{\mathrm{d}t^*}
-
\dfrac{12}{19}\hat{H}_{2,2}^*
\\*
+
\dfrac{1}{38}
\left\{
 \left[
  19\left(1+2\epsiloni^*\right)\vec{k}
  -
  2\left(11+19\epsiloni^*\right)\vec{D}_{1,0}^*
 \right]
 \vec{D}_{1,1}^*
\right\}^\mathrm{S}
\\*
-
\dfrac{1}{114}
 \left[
  19\left(1+2\epsiloni^*\right)\vec{k}
  -
  2\left(11+19\epsiloni^*\right)\vec{D}_{1,0}^*
 \right]
\cdot\vec{D}_{1,1}^*\;\hat{I}
\\*
-
\dfrac{2}{133}
\left[
 2\left(8-\epsiloni^*\right)\vec{k}
 -
 5\left(11-13\epsiloni^*\right)\vec{D}_{1,0}^*
\right]
\cdot\hat{D}_{3,1}^*
,
\label{B22}
\end{multline}
\begin{multline}
\hat{B}_{4,2}^*
=
-
\dfrac{12}{17}\hat{H}_{4,2}^*
\\
+
\dfrac{1}{102}
\left\{
 \left[
  17\left(1+6\epsiloni^*\right)\vec{k}
  -
  2\left(43+51\epsiloni^*\right)\vec{D}_{1,0}^*
 \right]
 \;\hat{D}_{3,1}^*
\right\}^\mathrm{S}
\\*
-
\dfrac{3}{714}
\left\{
 \left[
  17\left(1+6\epsiloni^*\right)\vec{k}
  -
  2\left(43+51\epsiloni^*\right)\vec{D}_{1,0}^*
 \right]
 \cdot\hat{D}_{3,1}^*\;\hat{I}
\right\}^\mathrm{S}
,
\label{B42}
\end{multline}
\begin{equation}
\hat\Omega_{3,2}^*
=
-
\dfrac{1}{10}
\left\{
 \left[
  \left(5 + 2\epsiloni^*\right)\vec{k}
  -
  2\left(1 + \epsiloni^*\right)\vec{D}_{1,0}^*
 \right]
 \times\hat{D}_{3,1}^*
\right\}^\mathrm{S}
.
\label{Omega32}
\end{equation}

The linear ordinary differential equations for $\hat{H}_{2,2}^*$ and $\hat{H}_{4,2}^*$, Eqs.~(\ref{ode:H22}) and  (\ref{ode:H42}), have the following solutions:
\begin{multline}
\hat{H}_{2,2}^*(t^*)
=
\exp\left(-\dfrac{40}{19}t^*\right)
\hat{H}_{2,2}^*(t^*_0)
\\
+
\exp\left(-\dfrac{40}{19}{t^*}\right)
\int\limits_{t^*_0}^{t^*}
 \exp\left(\dfrac{40}{19}{t^*}'\right)
 \hat{F}_{2,2}^*({t^*}')
\;\mathrm{d}{t^*}'
,
\label{H22}
\end{multline}
\begin{multline}
\hat{F}_{2,2}^*
=
-
\dfrac{15}{19}\hat{B}_{2,1}^*
+
\dfrac{136}{19}\hat{C}_{2,1}^*
\\*
+
\dfrac{1}{19}
\left\{
 \left[
  19\left(1+2\epsiloni^*\right)\vec{k}
  -
  2\left(7+19\epsiloni^*\right)\vec{D}_{1,0}^*
 \right]
 \vec{D}_{1,1}^*
\right\}^\mathrm{S}
\\*
-
\dfrac{1}{57}
 \left[
  19\left(1+2\epsiloni^*\right)\vec{k}
  -
  2\left(7+19\epsiloni^*\right)\vec{D}_{1,0}^*
 \right]
\cdot\vec{D}_{1,1}^*\;\hat{I}
\\*
+
\dfrac{12}{133}
\left[
 \left(8-\epsiloni^*\right)\vec{k}
 +
 \left(39-34\epsiloni^*\right)\vec{D}_{1,0}^*
\right]
\cdot\hat{D}_{3,1}^*
,
\label{F22}
\end{multline}
\begin{multline}
\hat{H}_{4,2}^*(t^*)
=
\exp\left(-\dfrac{72}{17}t^*\right)
\hat{H}_{2,2}^*(t^*_0)
\\
+
\int\limits_{t^*_0}^{t^*}
 \exp\left[\dfrac{72}{17}\left({t^*}'-{t^*}\right)\right]
 \hat{F}_{4,2}^*({t^*}')
\;\mathrm{d}{t^*}'
,
\label{H42}
\end{multline}
\begin{multline}
\hat{F}_{4,2}^*
=
\dfrac{1}{51}
\left\{
 \left[
  17\left(1+6\epsiloni^*\right)\vec{k}
  +
  2\left(41-51\epsiloni^*\right)\vec{D}_{1,0}^*
 \right]
 \;\hat{D}_{3,1}^*
\right\}^\mathrm{S}
\\*
-
\dfrac{1}{119}
\left\{
 \left[
  17\left(1+6\epsiloni^*\right)\vec{k}
  +
  2\left(41-51\epsiloni^*\right)\vec{D}_{1,0}^*
 \right]
 \cdot\hat{D}_{3,1}^*\;\hat{I}
\right\}^\mathrm{S}
.
\label{F42}
\end{multline}

The corrections to the stationary solutions are as follows:
\begin{equation}
\vec{D}_{1,1\mathrm{sr}}^*
=
{D}_{1,1\mathrm{sr}x}^*\vec{i}_x
+
{D}_{1,1\mathrm{sr}z}^*\vec{k}
,
\label{D11sr}
\end{equation}
\begin{multline}
{D}_{1,1\mathrm{sr}x}^*
=
-
\dfrac
 {3\left(2+\lambda_1^*\right)^3}
 {100\left(2+\epsiloni^*\right)
  \tau^{*2}
  \sqrt{
   \dfrac{3}{2}
   \dfrac
    {\tau^*\left(\epsiloni^*-\lambda_1^*\right)}
    {\left(2+\lambda_1^*\right)^2}
   -
   1
  }
 }
\\
\times
\left[
 \dfrac
  {9\left(\epsiloni^*-\lambda_1^*\right)}
  {\left(2+\lambda_1^*\right)^3}
 \tau^2
 +
 \dfrac
  {2\left(9\lambda_1^*+8\epsiloni^*+34\right)}
  {\left(2+\lambda_1^*\right)^2}
 \tau^*
 +
 32
\right]
,
\end{multline}
\begin{multline}
{D}_{1,1\mathrm{sr}z}^*
=
-
\dfrac
 {6\left(\epsiloni^*-\lambda_1^*\right)}
 {25\left(2+\epsiloni^*\right)\tau^{*2}}
\\
\times
\left[
 \left(\epsiloni^*+2\right)\tau^*
 +
 2\left(2+\lambda_1^*\right)^2
\right]
,
\end{multline}
\begin{equation}
\vec\Omega_{1,1\mathrm{sr}}^*
=
\Omega_{1,1\mathrm{sr}}^*\vec{i}_y
,
\end{equation}
\begin{multline}
\Omega_{1,1\mathrm{sr}}^*
=
\dfrac
 {3\left(2+\lambda_1^*\right)^3}
 {200\left(2+\epsiloni^*\right)
  \tau^{*2}
  \sqrt{
   \dfrac{3}{2}
   \dfrac
    {\tau^*\left(\epsiloni^*-\lambda_1^*\right)}
    {\left(2+\lambda_1^*\right)^2}
   -
   1
  }
 }
\\
\times
\left[
 \dfrac
  {9\left(\epsiloni^*-\lambda_1^*\right)}
  {\left(2+\lambda_1^*\right)^3}
 \tau^2
 +
 \dfrac
  {2\left(9\lambda_1^*+8\epsiloni^*+34\right)}
  {\left(2+\lambda_1^*\right)^2}
 \tau^*
 +
 32
\right]
.
\end{multline}
The stationary solution Eq. (\ref{ode:D31}) is
\begin{multline}
\hat{D}_{3,1\mathrm{sr}}^*
=
\dfrac
 {4\tau^*\left(4+3\lambdai^*+12\lambdas^*\right)
  \hat{G}_{3,1\mathrm{sr}}^*
 }
 {\left(4+3\lambdai^*+12\lambdas^*\right)^2
  +
  9\tau^{*2}\left(4 + 3\epsiloni^*\right)^2
  \Omega_{1,0\mathrm{sr}}^{*2}
 }
\\
+
\dfrac
 {12\tau^{*2}\left(4 + 3\epsiloni^*\right)
  \left(
   \vec\Omega_{1,0\mathrm{sr}}^*\times
   \hat{G}_{3,1\mathrm{sr}}^*
  \right)^\mathrm{S}
 }
 {\left(4+3\lambdai^*+12\lambdas^*\right)^2
  +
  9\tau^{*2}\left(4 + 3\epsiloni^*\right)^2
  \Omega_{1,0\mathrm{sr}}^{*2}
 }
,
\label{D31sr}
\end{multline}
\begin{multline}
\hat{G}_{3,1\mathrm{sr}}^*
=
  \left\{
   \left[
   \left(1 - \epsiloni^*\right) \vec{k}
   +
   \left(2 + \epsiloni^*\right) \vec{D}_{1,0\mathrm{sr}}^* 
   \right]\;
   \hat{B}_{2,1\mathrm{sr}}^*
  \right\}^\mathrm{S}
\\
-
\dfrac{2}{5}
  \left\{
   \left[
    \left(1 - \epsiloni^*\right) \vec{k}
    +
    \left(2 + \epsiloni^*\right) \vec{D}_{1,0\mathrm{sr}}^* 
   \right]\cdot
   \hat{B}_{2,1\mathrm{sr}}^*\;\hat{I}
  \right\}^\mathrm{S}
.
\label{G31sr}
\end{multline}

\end{document}